\documentclass[journal]{IEEEtran}

\usepackage{cite}
\usepackage{amsmath,amssymb,amsfonts}
\usepackage{graphicx}
\usepackage{textcomp}
\usepackage{xcolor}
\usepackage{booktabs}
\usepackage{multirow}
\usepackage{algorithm}
\usepackage{algpseudocode}
\usepackage{url}
\usepackage{hyperref}
\usepackage{array}
\usepackage{subcaption}
\usepackage{bm}
\usepackage{tikz}
\usepackage{xspace}
\usepackage{caption}
\usetikzlibrary{positioning,arrows.meta,fit,backgrounds,calc}

\makeatletter
\renewcommand\section{\@startsection{section}{1}{\z@}%
  {1.6ex plus 0.4ex minus 0.4ex}%
  {0.5ex plus 0.2ex minus 0ex}%
  {\normalfont\normalsize\centering\scshape}}
\renewcommand\subsection{\@startsection{subsection}{2}{\z@}%
  {1.6ex plus 0.4ex minus 0.4ex}%
  {0.4ex plus 0.2ex minus 0ex}%
  {\normalfont\normalsize\itshape}}
\makeatother
\begin{document}

\bstctlcite{IEEEexample:BSTcontrol}
\setlength{\abovedisplayskip}{3pt plus 1pt minus 1pt}
\setlength{\belowdisplayskip}{3pt plus 1pt minus 1pt}
\setlength{\abovedisplayshortskip}{1pt plus 1pt}
\setlength{\belowdisplayshortskip}{2pt plus 1pt minus 1pt}
\setlength{\textfloatsep}{6pt plus 2pt minus 2pt}
\setlength{\floatsep}{6pt plus 2pt minus 2pt}
\setlength{\intextsep}{6pt plus 2pt minus 2pt}

\renewcommand{\topfraction}{0.92}
\renewcommand{\bottomfraction}{0.85}
\renewcommand{\textfraction}{0.07}
\renewcommand{\floatpagefraction}{0.80}
\renewcommand{\dbltopfraction}{0.92}
\renewcommand{\dblfloatpagefraction}{0.80}
\setcounter{topnumber}{3}
\setcounter{bottomnumber}{2}
\setcounter{totalnumber}{4}
\setcounter{dbltopnumber}{3}

\title{A QoE-Driven Framework for O-RAN Network Slicing Based on Empirical Multi-User VR Traffic Modelling}

\author{
  Ahmed Chyad\textsuperscript{$\star$}
  \quad David Naseh
  \quad Maryam Hafeez
  \quad Syed Ali Raza Zaidi
  \\
  School of Electronic and Electrical Engineering, University of Leeds, Leeds, UK\\
}

\markboth{IEEE Transactions on XXXXXX, Vol.~XX, No.~X, MONTH~2026}%
{Chyad \MakeLowercase{\textit{et al.}}: A QoE-Driven Framework for O-RAN Network Slicing Based on Empirical Multi-User VR Traffic Modelling}

\maketitle

\begin{abstract}
Virtual Reality (VR) requires strict Quality-of-Experience (QoE) guarantees, making
radio resource management harder. Static slice allocations are wasteful, while dynamic
approaches often rely on heuristic penalty functions lacking an empirical basis. We
introduce a QoE-driven network-slicing framework for Open RAN in which resource
allocation is guided by empirically derived multi-user VR traffic statistics. Closed-form
$n$-fold convolutions model the aggregate data rate and inter-arrival time of $n$
concurrent users, validated empirically for up to four. The fitted data-rate
cumulative distribution function is checked against the measured throughput to give a
percentile-based satisfaction indicator; the inter-arrival law sets the slice
targets. A near-real-time RIC xApp closes a loop over the E2SM-KPM and E2SM-RC
interfaces to apply slice-level Physical Resource Block (PRB) quotas. On an
over-the-air testbed it reclaims up to 41\% of the cell's committed PRBs from a greedy
neighbouring slice while never reducing the protected VR slice. An analytical study then
estimates the PRB demand and per-user QoE across user counts and channel qualities,
giving the admission boundary beyond which the target QoE is unattainable. A controlled
comparison further shows that a fixed-rate synthetic load overstates the VR slice's PRB
footprint by 68\%, showing the necessity of data-driven slice dimensioning.
\end{abstract}



\section{Introduction}
\IEEEPARstart{V}{irtual} Reality (VR) services impose stringent network requirements due to the high sensitivity of immersive applications to network-induced degradations. End-to-end latency exceeding 20~ms can cause motion sickness, jitter beyond 5~ms can induce perceptible video and audio distortion, and packet loss above $1\%$ can severely degrade the user experience~\cite{ieee3079}. Unlike traditional multimedia streaming, where buffering can mask network variations, VR services demand real-time, bidirectional interaction with strict end-to-end latency budgets. Consequently, the network must sustain not only the required data rate but also the temporal regularity of VR traffic. Existing approaches to QoE-aware resource management for VR leave three gaps that this paper addresses.

\textbf{Heuristic, traffic-agnostic QoE mapping.} Many existing QoE models turn
network metrics into a satisfaction score through penalty functions, commonly
exponential, logarithmic, or linear, whose shape and parameters are typically selected
\emph{a priori} rather than derived from the statistics of the measured
traffic~\cite{ruan2021survey}. Data-driven
predictors increase accuracy but act as black boxes that neither expose the
traffic-statistical meaning of the score nor generalise beyond their training
distribution~\cite{kumar2021qoe,kougioumtzidis2023deep}. In neither case can the score be
seen as the likelihood that the current condition is acceptable relative to the actual VR
traffic.

\textbf{Provisioning from synthetic or single-user traffic.} Slice dimensioning and QoE
targets are frequently derived from fixed-rate sources or the standardised
truncated-Gaussian frame model, which underestimates the effective bandwidth of real VR
flows~\cite{korneev2024cloudvr} and, being single-user, offer no analytical basis for
sizing a slice that must handle several concurrent XR users. The aggregate demand of $n$
simultaneous users is thus left uncharacterised.

\textbf{Validation in simulation rather than on physical testbed.} QoE-driven XR slicing
has largely been tested in system-level simulation or with synthetically generated
traffic over emulated or wired links~\cite{paymard2023performance,casparsen2024near},
leaving unverified whether the same control behaves correctly under real link adaptation,
channel variation, and scheduler dynamics on an over-the-air O-RAN setup.

Left unaddressed, these gaps lead either to over-provisioning, which wastes radio
resources, or to under-provisioning, which degrades the immersive experience.
In order to address these gaps, this paper presents a QoE-driven network-slicing framework for VR services that derives
both its quality model and its resource-control parameters from empirically-fitted
VR-traffic distributions. Our key contributions are:

\begin{enumerate}
    \item \textbf{Empirical multi-user VR traffic model.} Extending our single-UE
    study~\cite{chyad2025modelling}, we derive closed-form $n$-fold convolutions for the
    aggregate inter-arrival time and data rate of $n$ concurrent VR users, validated by
    fitting across one to four users.

    \item \textbf{Percentile-based QoE model.} We evaluate the fitted data-rate CDF at the
    measured downlink throughput, ranking live performance against the traffic distribution
    on a mean-opinion-score (MOS) scale, and apply it in closed form to the per-user QoE
    across user counts and channel qualities.

    \item \textbf{Analytical multi-user capacity.} We express the PRBs required to serve
    $n$ users at the target QoE as a function of user count and wideband CQI, producing the
    admission boundary beyond which the aggregate demand exceeds a cell's PRB resources.

    \item \textbf{Closed-loop O-RAN PRB control.} We implement a near-real-time RIC xApp
    that closes a loop over E2SM-KPM and E2SM-RC on an OCUDU testbed, estimating per-PRB
    efficiency through a self-calibrating chain and setting priority-ordered slice quotas without per-packet capture at runtime.

    \item \textbf{Hardware validation with realistic traffic.} The controller recovers up
    to 41\% of the cell's allocated PRBs from a greedy neighbour while never reducing the
    protected VR-eMBB slice and holding loss within the IEEE~3079-2020 budget (0.037\%); a
    controlled comparison shows a fixed-rate synthetic load overstates the slice's PRB
    footprint by 68\%.
\end{enumerate}


\section{Related Work}
\label{sec:related}

\subsection{VR Quality of Experience}
Quality of Experience measures service quality as perceived by the end user, and for
immersive media it is a multidimensional quantity influenced by coding, network transport and
human perceptual factors. Surveys of VR and 360-degree video establish QoE as the
appropriate optimisation target, while observing that metrics designed for planar video
correlate only loosely with the perceived quality of immersive
content~\cite{ruan2021survey,chiariotti2021survey}. What they do not offer is a
lightweight, traffic-grounded measure suitable for real-time in-network control. A
prominent line of work instead predicts QoE directly from network measurements using
machine learning~\cite{kumar2021qoe}. Closest to the present setting, Kougioumtzidis et
al.~\cite{kougioumtzidis2023deep} estimate the transmission-related QoE of VR 360-degree
video over Open RAN in real time, using an encoder--decoder long short-term memory network
driven by measurable QoS parameters. Such learned predictors are accurate, but they
operate as black boxes: they map QoS to a QoE value without exposing the
traffic-statistical basis of the score, and they estimate quality rather than deriving the
resource-control targets needed to sustain it. The present work, instead, uses an interpretable, empirically grounded measure, namely, the fitted VR data-rate CDF evaluated at the measured downlink throughput, and couples it to a closed-loop O-RAN controller whose service-rate and scheduling-gap targets derive from that same fitted model.

\subsection{Network Slicing for VR}
Network slicing combines heterogeneous services on shared infrastructure, and its
enabling technologies, software-defined networking and network-function virtualisation,
have been surveyed in depth~\cite{zhang2017network,condoluci2017enabling}. These works
define the slicing foundation but not the runtime allocation policy within a slice, which
in conventional deployments is static or coarsely provisioned. Committed radio resources
are then wasted whenever the channel is good or the source is idle, which motivates
the fine-grained, traffic-driven reallocation pursued here.

Most directly comparable is the work of Casparsen et al.~\cite{casparsen2024near}, whose
near-RT RIC xApp guides the experienced latency of a VR stream toward a requested value
while releasing RAN resources for a secondary user, reporting $15.8\%$ more resources to
that user than a latency-equivalent static allocation. That study shares our objective but differs
in two respects that make the present evaluation more realistic of over-the-air
operation: it drives a 4G link with synthetically generated traffic over a wired
UE-to-gNB connection, whereas our controller runs on a 5G~NR air interface with
empirically-collected VR traffic, so that real link adaptation and channel variation are
exercised rather than emulated; and its control law is a heuristic latency regulator,
whereas ours derives its service-rate and scheduling-gap targets in closed form from the
fitted traffic distributions.

\subsection{VR Traffic Modelling}
Measurement-driven models match parametric laws to captured gameplay. Zhao et
al.~\cite{zhao2021virtual} fit frame size to a log-logistic distribution and frame
inter-arrival time to a Burr distribution on a cloud-VR testbed, and Liubogoshchev et
al.~\cite{liubogoshchev2021adaptive} develop an analytical model and optimisation
framework for adaptive cloud-based XR. Korneev et al.~\cite{korneev2024cloudvr} extend
this line and show that the 3GPP baseline model underestimates the effective VR
bitrate, the bandwidth actually required to deliver the stream by roughly $25\%$. Our
measurements independently back up that conclusion: at 120~Hz the captured trace
sustains a measured mean downlink rate of 151.5~Mbps and peaks at 250~Mbps, well above the
compressed-media figures assumed by simplified baseline models. At the standards level, the 3GPP statistical XR model~\cite{3gpp38838,3gpp26926} represents frame arrivals as
quasi-periodic with truncated-Gaussian jitter, a form intended for simulation
tractability rather than packet-level fidelity.

All of these models are single-user. Building on our own single-UE
model~\cite{chyad2025modelling}, this work derives closed-form $n$-fold convolution
expressions for the aggregate inter-arrival time and data rate of $n$ concurrent VR users
(Section~\ref{sec:multiuser-convolution}) and validates them empirically.

\section{VR Traffic Modelling and Multi-User Aggregation}
\label{sec:multiuser-convolution}
\label{sec:traffic-char}

\subsection{Single-User Distribution Fitting}
\label{sec:Distribution-Fitting}
The testbed and the sample applications are those of our previous
work~\cite{chyad2025modelling}. Candidate distributions for both the packet inter-arrival
time (IAT) and the data rate are ranked by the Kullback--Leibler ($D_{\mathrm{KL}}$)
divergence, which measures how one distribution $P$ diverges from a reference
$Q$~\cite{KullbackLeibler1951},
with $D_{\mathrm{KL}}=0$ for identical distributions. The captured VR traffic has a mean
packet size $S_{\mathrm{pkt}}=1080.3$~bytes, and its IATs reflect the burst structure of
encoded VR: each frame is sent as a dense burst of back-to-back packets separated by an
idle gap. Within a burst, arrivals are microseconds apart, whereas the inter-burst gaps are
three orders of magnitude larger, so each burst carries about $110$ packets
($\approx119$~kB). The dense within-burst mode determines how the RAN must serve the flow,
since that is where the packets requiring transmission arrive; the \emph{unconditional}
mean IAT $\mathbb{E}[X]_{\mathrm{unc}}=57.04~\mu$s, which integrates the idle gaps,
corresponds to a measured mean wire rate of
$8S_{\mathrm{pkt}}/\mathbb{E}[X]_{\mathrm{unc}}=151.5$~Mbps.

The packet IAT is modelled by the \textbf{Generalized Extreme Value (GEV)} distribution,
which gave the lowest $D_{\mathrm{KL}}$. Writing $z=(x-\mu_X)/\sigma_X$, its density
is~\cite{kotz2000extreme}
\begin{equation}
f(x) = \begin{cases}
\dfrac{1}{\sigma_X}\left(1+\xi z\right)^{-1/\xi-1}
\exp\!\left\{-\left(1+\xi z\right)^{-1/\xi}\right\}, & \xi \neq 0,\\[6pt]
\dfrac{1}{\sigma_X}\,e^{-z}\exp\!\left\{-e^{-z}\right\}, & \xi = 0,
\end{cases}
\label{eq:gev}
\vspace{-0.2cm}
\end{equation}
with shape $\xi$, location $\mu_X$ and scale $\sigma_X>0$ (Table~\ref{tab:conv-constants}).
The fitted shape is negative, placing the fit in the bounded Weibull (Type~III) domain,
consistent with VR frame-timing constraints: the within-burst law ends at the finite
upper endpoint $h_X=\mu_X-\sigma_X/\xi=9.727~\mu$s, which is the support carried into the
convolution below. This within-burst support must not be interchanged with
$\mathbb{E}[X]_{\mathrm{unc}}$, which is an order of magnitude larger because it includes
the inter-burst gaps.

The data rate is asymmetric, with a heavier left tail, and is best fitted by the
\textbf{Generalized Logistic (GL)} distribution, whose density is~\cite{abberger2002ml}
\begin{equation}
f(y) = \frac{b}{\sigma_Y}\,
\frac{e^{-(y-\theta)/\sigma_Y}}{\left(1 + e^{-(y-\theta)/\sigma_Y}\right)^{b+1}},
\quad b,\sigma_Y > 0,\ \theta \in \mathbb{R},
\label{eq:genlogistic}
\end{equation}
with location $\theta$, scale $\sigma_Y$ and shape $b$ (Table~\ref{tab:conv-constants}).
All fitted parameters and the derived convolution constants are listed in
Table~\ref{tab:conv-constants}.

\subsection{Aggregation Model}
A fully coupled model would treat each user's behaviour as dependent on the instantaneous
radio state (CQI, SINR, MCS, PRB allocation, numerology, bandwidth, and carrier), which is
computationally intractable. We therefore use the standard assumption that concurrent VR
users are independent and identically distributed (i.i.d.), so that the aggregate traffic
of $n$ users is the convolution of the single-user densities, producing closed-form expressions for capacity planning.

For $n$ i.i.d.\ users, the aggregate IAT $S_n$ and aggregate data rate $R_n$ are the sums
\begin{equation}
S_n=\sum_{i=1}^{n}X_i,\qquad R_n=\sum_{i=1}^{n}Y_i,
\label{eq:sums}
\end{equation}
with $X_i\sim\mathrm{GEV}(\xi,\mu_X,\sigma_X)$ and $Y_i\sim\mathrm{GL}(b,\theta,\sigma_Y)$.
The density $f_{\Sigma_n}$ of a sum of independent variables is the $n$-fold convolution
of the common density $g$~\cite{feller1966introduction}, evaluated below once as a generic
bounded-support convolution and then specialised to the GEV and GL cases.


\begin{table}[!t]
\caption{Fitted single-user VR traffic parameters and derived convolution constants.}
\label{tab:conv-constants}
\centering
\footnotesize
\begin{tabular}{lll}
\toprule
\textbf{Symbol} & \textbf{Definition} & \textbf{Value}\\
\midrule
\multicolumn{3}{l}{\emph{GEV inter-arrival time (ms)}}\\
$\xi$        & shape                              & $-0.383505$\\
$\mu_X$      & location                           & $5.210\times10^{-3}$\\
$\sigma_X$   & scale                              & $1.732\times10^{-3}$\\
$\ell_X$     & lower support (empirical)          & $1.000\times10^{-3}$\\
$h_X$        & upper endpoint $\mu_X-\sigma_X/\xi$ & $9.727\times10^{-3}$\\
$\kappa_X$   & amplitude $1/\sigma_X$             & $577.37$\\
$\lambda_X$  & slope $-\xi/\sigma_X$              & $221.42$\\
$\eta$       & tail index $-1/\xi$                & $2.6075$\\
\midrule
\multicolumn{3}{l}{\emph{Generalized Logistic data rate (Mbps)}}\\
$b$          & shape                              & $0.55$\\
$\theta$     & location                           & $173.54$\\
$\sigma_Y$   & scale                              & $14.81$\\
$\ell_Y$     & lower support                      & $70$\\
$h_Y$        & upper support                      & $250$\\
$\kappa_Y$   & amplitude $b/\sigma_Y$             & $0.0371$\\
$\beta$      & exponent $b+1$                     & $1.55$\\
$\zeta_{\mathrm{lo}}$ & $(\ell_Y-\theta)/\sigma_Y$ & $-6.99$\\
$\zeta_{\mathrm{hi}}$ & $(h_Y-\theta)/\sigma_Y$    & $5.16$\\
\bottomrule
\end{tabular}
\end{table}

\subsection{Generic Bounded-Support Convolution}
\label{sec:generic-conv}

Let $g$ be a density supported on $[\ell,h]$. Writing $t_n=s-\sum_{i=1}^{n-1}t_i$, the
$n$-fold convolution is the $(n-1)$-dimensional integral
\begin{equation}
f_{\Sigma_n}(s)=\idotsint_{\mathcal{D}_{n-1}(s)}\ \prod_{i=1}^{n}g(t_i)\;
dt_1\cdots dt_{n-1},
\label{eq:generic-conv}
\end{equation}
supported on $s\in[\,n\ell,\,nh\,]$, where the domain $\mathcal{D}_{n-1}(s)$ enforces
$t_i\in[\ell,h]$ for every $i$ (including $t_n$). The domain is given by the
sequential limits, for $j=1,\dots,n-1$,
\begin{equation}
\begin{aligned}
L_j&=\max\!\Big\{\ell,\ s-\textstyle\sum_{k=1}^{j-1}t_k-(n-j)\,h\Big\},\\[2pt]
U_j&=\min\!\Big\{h,\ s-\textstyle\sum_{k=1}^{j-1}t_k-(n-j)\,\ell\Big\},
\end{aligned}
\label{eq:generic-limits}
\end{equation}
so that $\int_{L_1}^{U_1}\!\cdots\!\int_{L_{n-1}}^{U_{n-1}}$
realises~\eqref{eq:generic-conv}. Equations~\eqref{eq:generic-conv}--\eqref{eq:generic-limits}
are used unchanged to both traffic quantities; only $g$ and $[\ell,h]$ differ. The
aggregate supports follow as $[n\ell_X,nh_X]$ and $[n\ell_Y,nh_Y]$, reaching
$[0.004,0.0389]$~ms and $[280,1000]$~Mbps at $n=4$.
 

%
\subsection{Inter-Arrival Time Aggregation (GEV)}
\label{sec:iat-conv}

In computational form, the single-user GEV density is
\begin{equation}
g_X(x)=\kappa_X\,[w(x)]^{\,\eta-1}\exp\!\big\{-[w(x)]^{\eta}\big\},
\label{eq:gev-comp}
\end{equation}
on $x\in[\ell_X,h_X]$, where the kernel $w(x)=1-\lambda_X(x-\mu_X)$ decreases from
$w(\ell_X)\!\approx\!1.93$ to $w(h_X)=0$ at the finite upper endpoint, and the exponents
$\eta-1=1.6075$ and $\eta=2.6075$ follow from $\xi$. Substituting $g\!=\!g_X$ and
$[\ell,h]\!=\![\ell_X,h_X]$ into~\eqref{eq:generic-conv} gives the aggregate IAT density
\begin{equation}
\begin{aligned}
f_{S_n}(s)=\kappa_X^{\,n}\!\idotsint_{\mathcal{D}_{n-1}(s)}\!
\prod_{i=1}^{n}[w(t_i)]^{\eta-1}\\
\times\exp\!\Big\{-\!\sum_{i=1}^{n}[w(t_i)]^{\eta}\Big\}\,d\mathbf{t},
\end{aligned}
\label{eq:gev-nfold}
\end{equation}
with $d\mathbf{t}=dt_1\cdots dt_{n-1}$, limits from~\eqref{eq:generic-limits} with
$(\ell,h)=(\ell_X,h_X)$, and support $s\in[\ell_X n,h_X n]$. As the base case, the
two-user density ($n=2$, $\kappa_X^2\!\approx\!3.33\times10^{5}$) is
\begin{equation}
\begin{aligned}
f_{S_2}(s)=\kappa_X^{2}\!\int_{L}^{U}\![w(x)]^{\eta-1}[w(s-x)]^{\eta-1}\\
\times\exp\!\big\{-[w(x)]^{\eta}-[w(s-x)]^{\eta}\big\}\,dx,
\end{aligned}
\label{eq:gev-n2}
\end{equation}
with $L=\max(\ell_X,\,s-h_X)$ and $U=\min(h_X,\,s-\ell_X)$. The exponential term does not
factor, so~\eqref{eq:gev-nfold} is evaluated numerically for $n\ge2$.

\subsection{Data-Rate Aggregation (Generalized Logistic)}
\label{sec:rate-conv}

The single-user GL density, with the standardised variable $z(y)=(y-\theta)/\sigma_Y$, is
\begin{equation}
g_Y(y)=\kappa_Y\,\frac{e^{-z(y)}}{\big(1+e^{-z(y)}\big)^{\beta}},\qquad
y\in[\ell_Y,h_Y].
\label{eq:gl-comp}
\end{equation}
Here the numerators telescope under summation: since
$\sum_{i=1}^{n}z(y_i)=(\,\textstyle\sum_i y_i-n\theta)/\sigma_Y=(r-n\theta)/\sigma_Y$,
defining
\begin{equation}
S_n^{\!*}(r)=\frac{r-n\theta}{\sigma_Y}
\label{eq:gl-Sn}
\end{equation}
gives $\prod_i e^{-z(y_i)}=e^{-S_n^{\!*}(r)}$, which factors out of the integral.
Substituting into~\eqref{eq:generic-conv} and changing variables $y_i\!\to\!z_i$
(each $dy_i=\sigma_Y\,dz_i$, so the $n-1$ free variables contribute a Jacobian factor
$\sigma_Y^{\,n-1}$) yields the aggregate data-rate density
\begin{equation}
\begin{aligned}
f_{R_n}(r)=\kappa_Y^{\,n}\,\sigma_Y^{\,n-1}\,e^{-S_n^{\!*}(r)}\\
\times\idotsint_{\mathcal{D}_{n-1}^{z}}\ \prod_{i=1}^{n}\big(1+e^{-z_i}\big)^{-\beta}\,
d\mathbf{z},
\end{aligned}
\label{eq:gl-nfold}
\end{equation}
with $z_n=S_n^{\!*}(r)-\sum_{i=1}^{n-1}z_i$, support $r\in[\ell_Y n,h_Y n]$, and limits
given by~\eqref{eq:generic-limits} evaluated in the $z$-domain, i.e.\ with
$(\ell,h,s)\!\to\!(\zeta_{\mathrm{lo}},\zeta_{\mathrm{hi}},S_n^{\!*}(r))$. The base case
($n=2$, $\kappa_Y^{2}\sigma_Y\!\approx\!0.0204$) is
\begin{equation}
\begin{aligned}
f_{R_2}(r)=\kappa_Y^{2}\sigma_Y\,e^{-S_2^{\!*}(r)}\!\int_{L^{z}}^{U^{z}}\!
\big(1+e^{-z}\big)^{-\beta}\\
\times\big(1+e^{-(S_2^{\!*}(r)-z)}\big)^{-\beta}\,dz,
\end{aligned}
\label{eq:gl-n2}
\end{equation}
with $L^{z}=\max(\zeta_{\mathrm{lo}},\,S_2^{\!*}(r)-\zeta_{\mathrm{hi}})$ and
$U^{z}=\min(\zeta_{\mathrm{hi}},\,S_2^{\!*}(r)-\zeta_{\mathrm{lo}})$. The closed-form
amplitude factor $\kappa_Y^{\,n}\sigma_Y^{\,n-1}e^{-S_n^{\!*}(r)}$ makes the GL aggregate
cheaper to evaluate than the GEV one, the remaining integral being a smooth product of
logistic kernels. The aggregate means scale linearly with the number of users, $\mathbb{E}[S_n]=n\,\mathbb{E}[X]$ and $\mathbb{E}[R_n]=n\,\mathbb{E}[Y]$ by linearity of expectation for i.i.d.\ summands~\cite{feller1966introduction}, while the aggregate supports $[n\ell,nh]$ give the extreme values it can attain, all $n$ users being simultaneously at an endpoint. Together, these bracket the per-cell rate envelope used for slice sizing in Section~\ref{sec:prb-analysis}, for which the $99$th percentile of the aggregate is the operative statistic rather than the endpoint itself.

\section{Multi-User Empirical Traffic Fitting}
\label{sec:multiuser-fit}
This section shows the results of statistical distribution fitting for VR traffic
captured across four UE-density scenarios (one to four VR UEs). Two traffic characteristics are modelled independently: the downlink data rate and the packet IAT. 


\subsection{Data-Rate Distribution Fitting Results}
Table~\ref{tab:top_five_distributions} presents the five best-fitting distributions for
the VR downlink data rate across the four UE scenarios. The GL distribution achieves rank~1 in all four cases, with $D_{\mathrm{KL}}$ values ranging from
0.0095 (one UE) to 0.0682 (two UEs), confirming its consistent superiority for
data-rate modelling.
 
\begin{table}[!t]
\caption{Top five fitted distributions for the VR data rate.}
\label{tab:top_five_distributions}
\centering
\footnotesize
\setlength{\tabcolsep}{2.5pt}
\renewcommand{\arraystretch}{1.05}
\begin{tabular}{@{}r@{\hskip 3pt}lc@{\hskip 7pt}r@{\hskip 3pt}lc@{}}
\toprule
\multicolumn{3}{c}{\textbf{One VR UE}} & \multicolumn{3}{c}{\textbf{Two VR UEs}}\\
\cmidrule(lr){1-3}\cmidrule(lr){4-6}
 & Distribution & $D_{\mathrm{KL}}$ & & Distribution & $D_{\mathrm{KL}}$\\
\midrule
1 & Generalized Logistic & 0.0095 & 1 & Generalized Logistic & 0.0682\\
2 & Generalized Gamma    & 0.0582 & 2 & Cauchy               & 0.0803\\
3 & Log-Gamma            & 0.0590 & 3 & Generalized Normal   & 0.0852\\
4 & Weibull              & 0.0591 & 4 & Laplace              & 0.0873\\
5 & Beta                 & 0.0601 & 5 & Expon.\ Weibull      & 0.0976\\
\midrule
\multicolumn{3}{c}{\textbf{Three VR UEs}} & \multicolumn{3}{c}{\textbf{Four VR UEs}}\\
\cmidrule(lr){1-3}\cmidrule(lr){4-6}
 & Distribution & $D_{\mathrm{KL}}$ & & Distribution & $D_{\mathrm{KL}}$\\
\midrule
1 & Generalized Logistic & 0.0665 & 1 & Generalized Logistic & 0.0625\\
2 & Cauchy               & 0.0805 & 2 & Cauchy               & 0.0762\\
3 & Generalized Normal   & 0.0853 & 3 & Generalized Normal   & 0.0815\\
4 & Laplace              & 0.0868 & 4 & Laplace              & 0.0842\\
5 & Log-Gamma            & 0.0956 & 5 & Gumbel               & 0.0888\\
\bottomrule
\end{tabular}
\end{table}
 
\begin{figure*}[t]
    \centering
    \begin{subfigure}{0.4\textwidth}
        \centering
        \includegraphics[width=\linewidth]{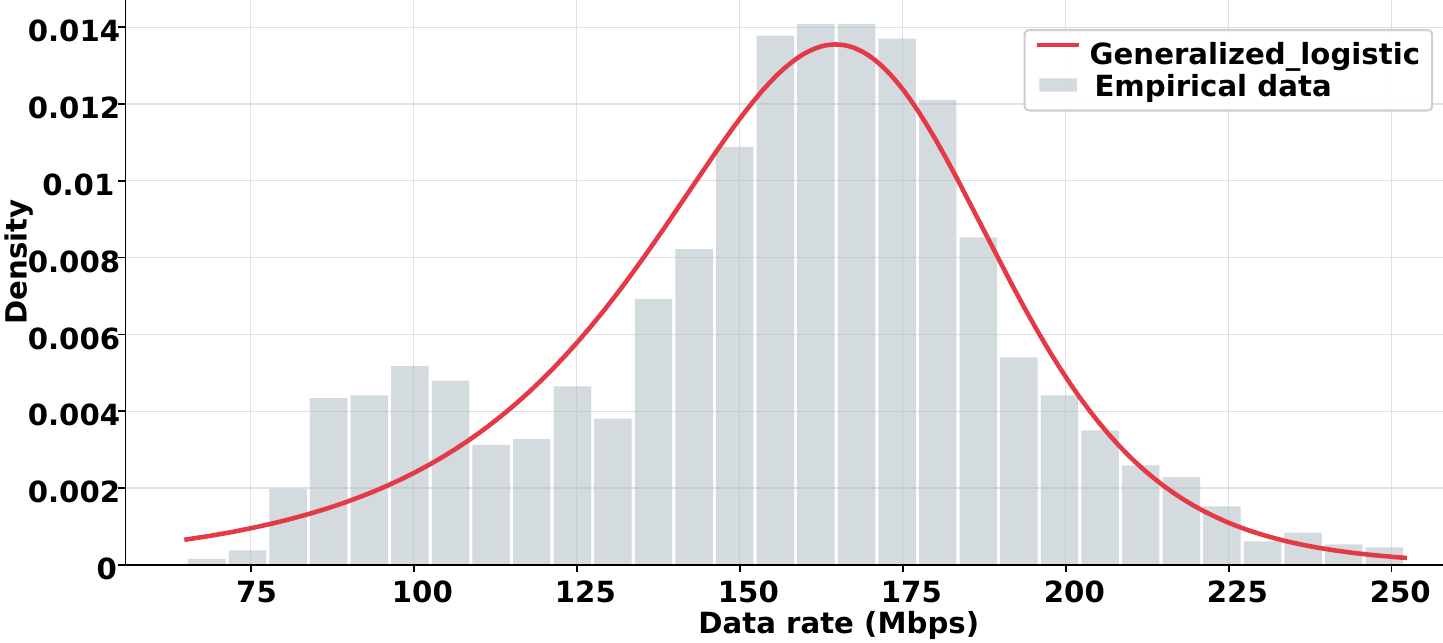}
        \caption{1 UE}
    \vspace{-0.1cm}     
    \end{subfigure}
    \hfill
    \begin{subfigure}{0.4\textwidth}
        \centering
        \includegraphics[width=\linewidth]{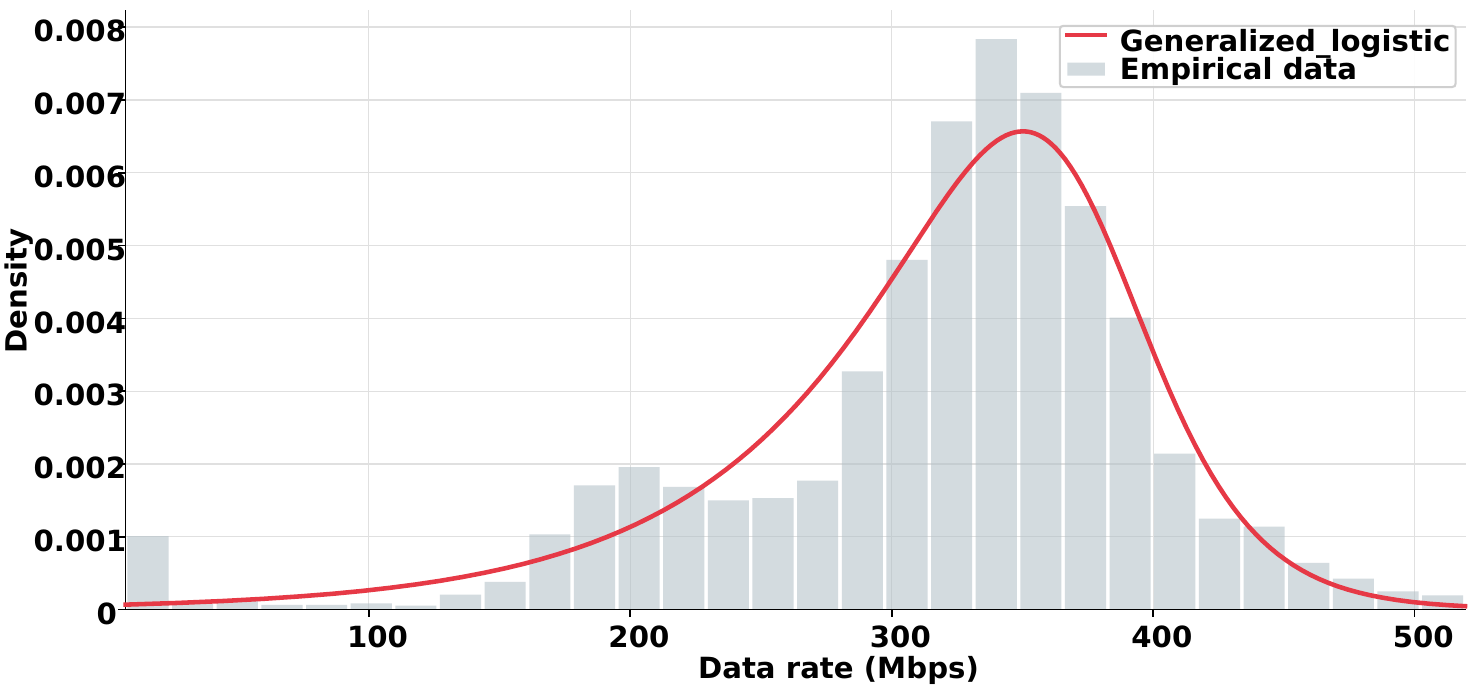}
        \caption{2 UEs}
    \vspace{-0.1cm}     
    \end{subfigure}
   
    \begin{subfigure}{0.4\textwidth}
        \centering
        \includegraphics[width=\linewidth]{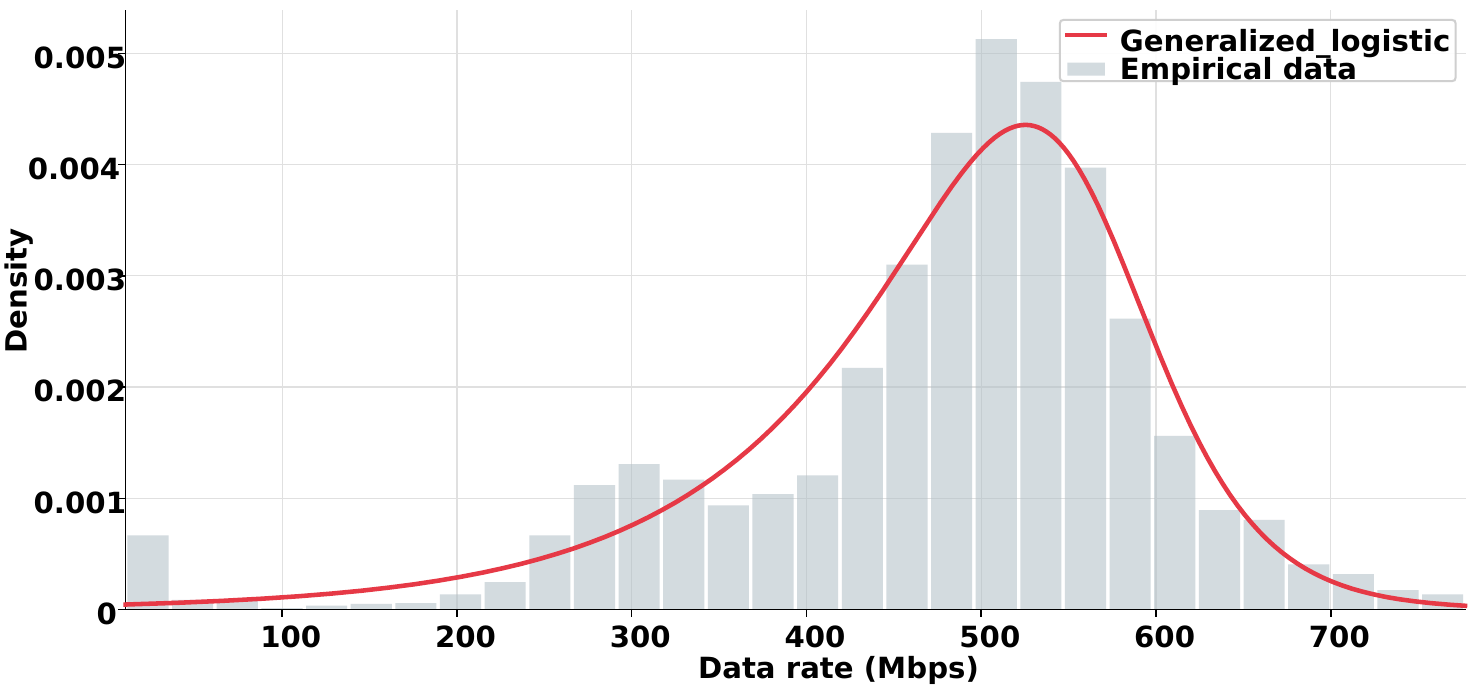}
        \caption{3 UEs}
    \vspace{-0.1cm}     
    \end{subfigure}
    \hfill
    \begin{subfigure}{0.4\textwidth}
        \centering
        \includegraphics[width=\linewidth]{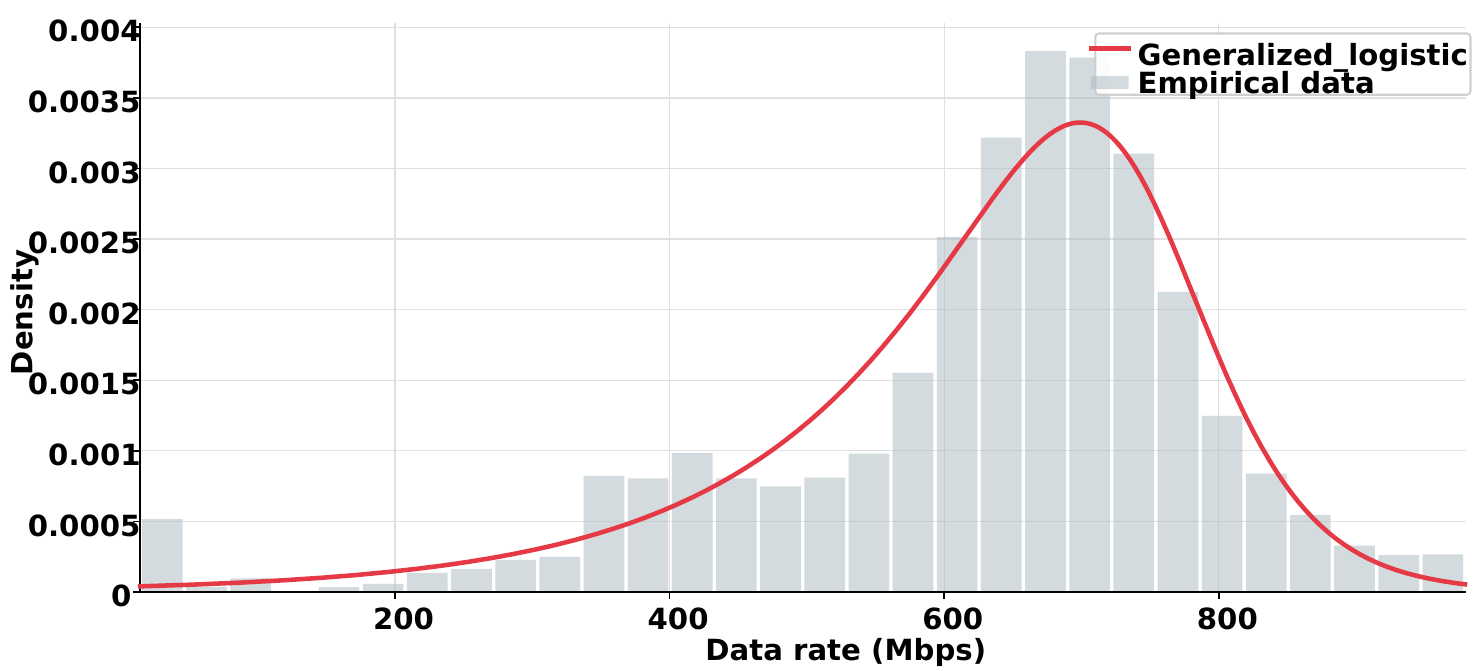}
        \caption{4 UEs}
    \vspace{-0.1cm}     
    \end{subfigure}
    \caption{Generalized Logistic fits to the VR downlink data rate for one to four
    concurrent VR UEs.}
    \label{fig:ue_comparison}
\vspace{-0.3cm}    
\end{figure*}

The consistency of the GL distribution across all UE densities, as shown in Fig.~\ref{fig:ue_comparison}, is
attributable to its symmetric, bell-shaped density with heavier tails than the
Gaussian, which naturally accommodates the moderate variability in aggregated VR
throughput arising from codec rate adaptation and scheduling fluctuations.
 
\subsection{Packet Inter-Arrival Time Distribution Fitting Results}
\begin{figure*}[t]
    \centering
    \begin{subfigure}{0.4\textwidth}
        \centering
        \includegraphics[width=\linewidth]{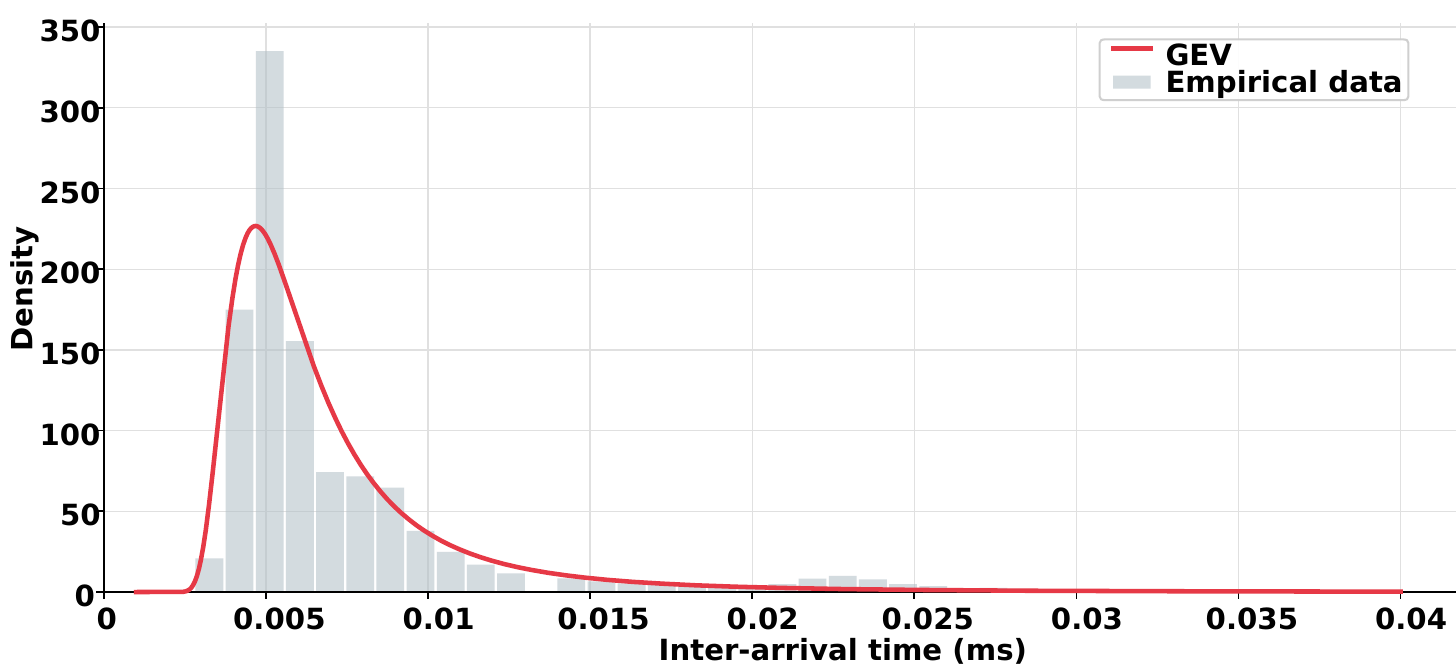}
        \caption{1 UE}
    \vspace{-0.1cm}    
    \end{subfigure}
    \hfill
    \begin{subfigure}{0.4\textwidth}
        \centering
        \includegraphics[width=\linewidth]{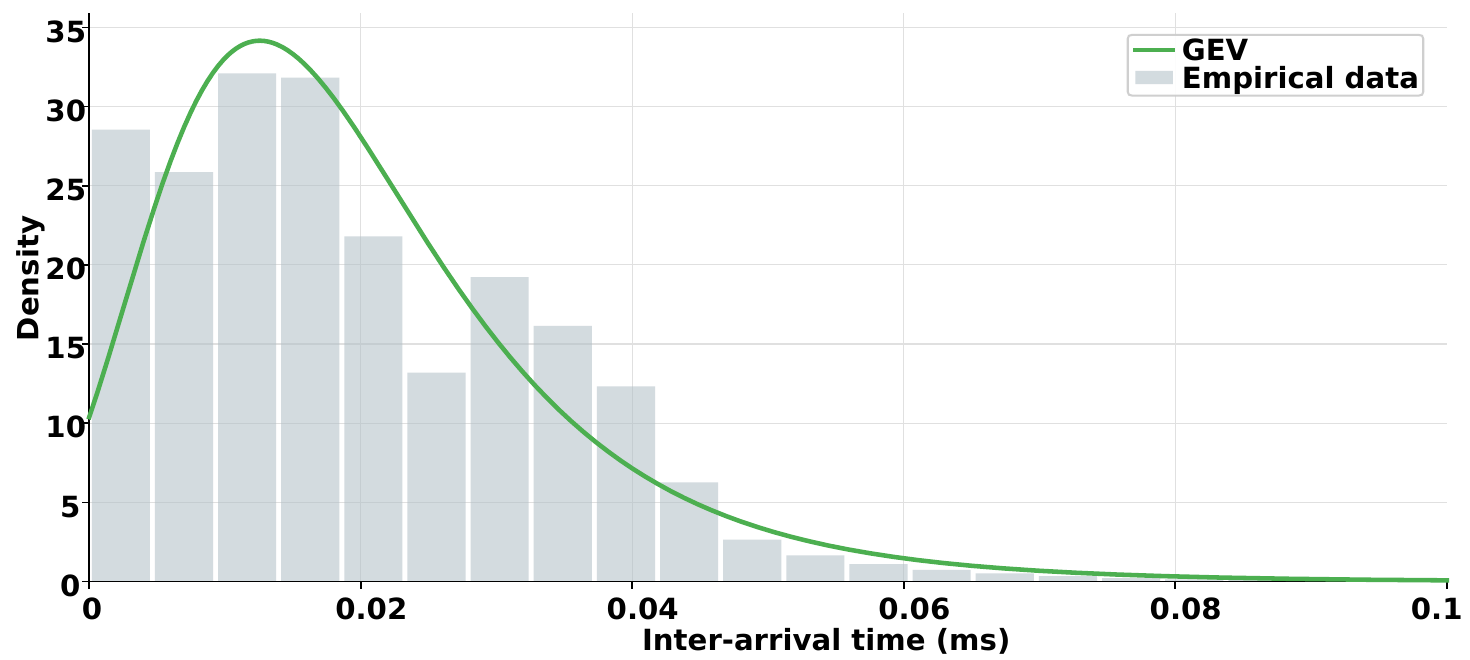}
        \caption{2 UEs}
    \vspace{-0.1cm}    
    \end{subfigure}
    
    \begin{subfigure}{0.4\textwidth}
        \centering
        \includegraphics[width=\linewidth]{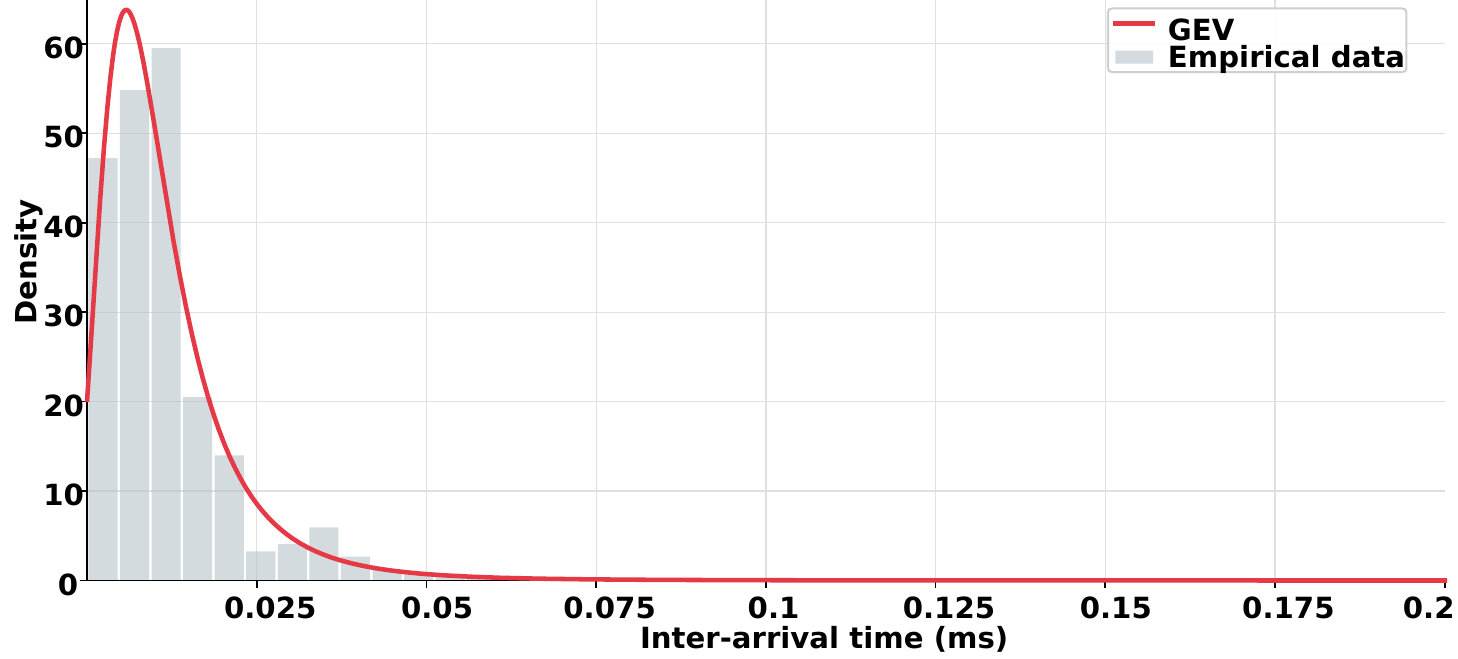}
        \caption{3 UEs}
    \vspace{-0.1cm}    
    \end{subfigure}
    \hfill
    \begin{subfigure}{0.4\textwidth}
        \centering
        \includegraphics[width=\linewidth]{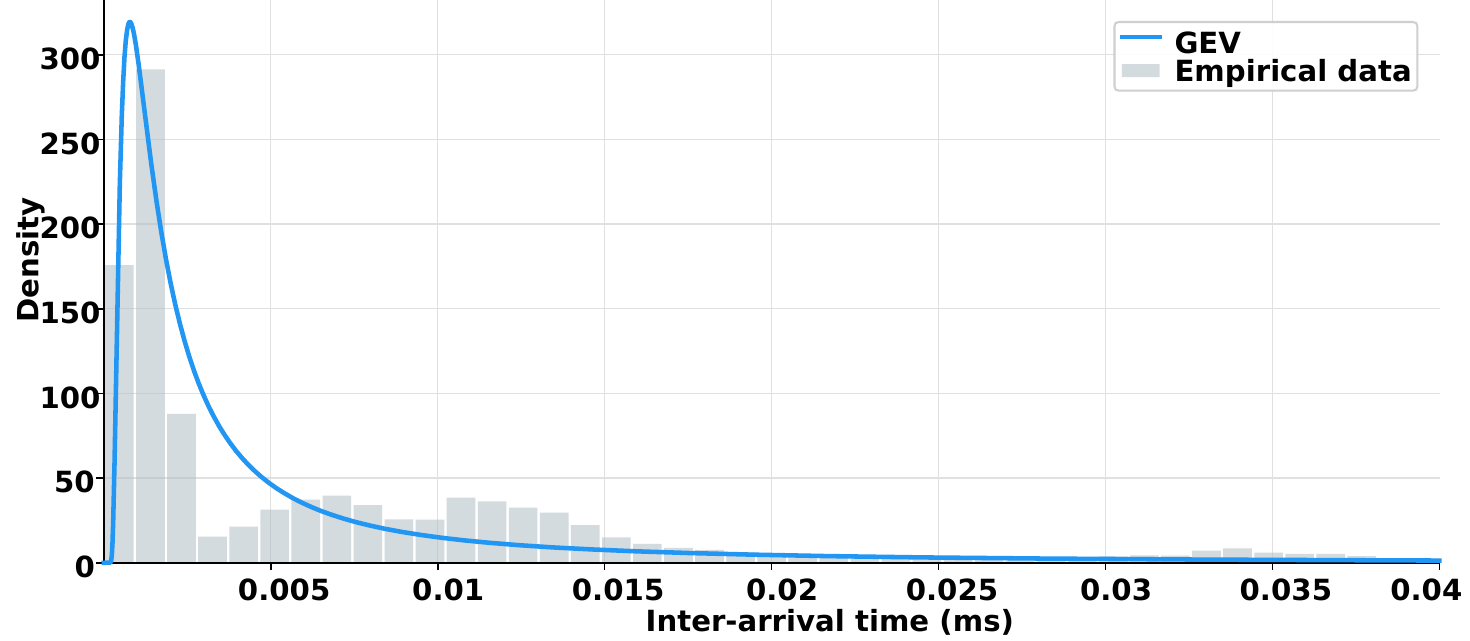}
        \caption{4 UEs}
    \vspace{-0.1cm}    
    \end{subfigure}
    \caption{GEV fits to the VR packet inter-arrival time for one to four concurrent
    VR UEs.}
    \label{fig:GEV_fitting}
    \vspace{-0.4cm}
\end{figure*}

Table~\ref{tab:top_five_interarrival} reports the ranking for the packet IAT, where
no single distribution achieves rank~1 in every scenario. The GEV and the Lognormal are the only two present in the top five at all four UE
densities, and of these only the GEV reaches rank~1, doing so for one UE, and three UEs; it is rank~4 for two UEs and rank~2 for four UEs. The corresponding fits are illustrated in
Fig.~\ref{fig:GEV_fitting}.

\begin{table}[!t]
\caption{Top five fitted distributions for the VR packet IAT.}
\label{tab:top_five_interarrival}
\centering
\footnotesize
\setlength{\tabcolsep}{2.5pt}
\renewcommand{\arraystretch}{1.05}
\begin{tabular}{@{}r@{\hskip 3pt}lc@{\hskip 7pt}r@{\hskip 3pt}lc@{}}
\toprule
\multicolumn{3}{c}{\textbf{One VR UE}} & \multicolumn{3}{c}{\textbf{Two VR UEs}}\\
\cmidrule(lr){1-3}\cmidrule(lr){4-6}
 & Distribution & $D_{\mathrm{KL}}$ & & Distribution & $D_{\mathrm{KL}}$\\
\midrule
1 & GEV                & 0.1112 & 1 & Weibull              & 0.0321\\
2 & Burr               & 0.1238 & 2 & Lognormal            & 0.0422\\
3 & Cauchy             & 0.1941 & 3 & Generalized Logistic & 0.0445\\
4 & Lognormal          & 0.2209 & 4 & GEV                  & 0.0454\\
5 & Generalized Pareto & 0.2950 & 5 & Burr                 & 0.0494\\
\midrule
\multicolumn{3}{c}{\textbf{Three VR UEs}} & \multicolumn{3}{c}{\textbf{Four VR UEs}}\\
\cmidrule(lr){1-3}\cmidrule(lr){4-6}
 & Distribution & $D_{\mathrm{KL}}$ & & Distribution & $D_{\mathrm{KL}}$\\
\midrule
1 & GEV                  & 0.0339 & 1 & L\'evy           & 0.1744\\
2 & Lognormal            & 0.0409 & 2 & GEV              & 0.1792\\
3 & Generalized Logistic & 0.0571 & 3 & Lognormal        & 0.1837\\
4 & Power Lognormal      & 0.0575 & 4 & Power Lognormal  & 0.1842\\
5 & Chi-Squared          & 0.0582 & 5 & Pareto           & 0.1951\\
\bottomrule
\end{tabular}
\end{table}
 
\subsection{Suitability of the GEV Model for VR IAT}
\label{sec:gev-tails}
The GEV is not an arbitrary curve-fitting choice, but follows from the structure of the
measurement. The Fisher--Tippett--Gnedenko theorem of Extreme Value Theory (EVT)
shows that normalised block maxima of i.i.d.\ variables converge to one of only
three limiting forms, Gumbel ($\xi=0$), Fr\'echet ($\xi>0$) or Weibull ($\xi<0$), unified
by the three-parameter form of~\eqref{eq:gev}~\cite{fisher1928limiting,coles2001extreme}.
That mechanism applies here because VR traffic is quasi-periodic at the codec frame rate,
yet the IAT observed at the gNB is perturbed by competing jitter sources: encoding-delay
variation, packetisation and segmentation, scheduling latency and transport-layer
retransmissions. Whichever source dominates at a given instant governs the realised IAT,
which is precisely the maximum-of-competing-effects structure EVT
models~\cite{liu2007evtwireless}.

The practical result is that the shape parameter $\xi$ adapts a single parametric
form to three tail regimes: heavy-tailed Fr\'echet behaviour for outlier gaps caused by
frame drops, deep fades or HARQ retransmissions; bounded Weibull support for packets that
exceed the delay budget of the QoS flow and are therefore discarded and never observed;
and light-tailed Gumbel decay for steady-state streaming. No competitor in
Table~\ref{tab:top_five_interarrival} covers this range. The Weibull cannot capture extreme
heavy tails~\cite{rinne2008weibull}, while the L\'evy, whose density decays as $x^{-3/2}$
with infinite mean and variance, is unsuitable for bounded or light-tailed
regimes~\cite{zolotarev1986stable}.

The GEV is also more flexible than the
3GPP truncated-Gaussian baseline of Section~\ref{sec:related}, which it includes as the
Gumbel limit $\xi\to0$; EVT has precedent in wireless traffic
analysis~\cite{liu2007evtwireless}, and GEV models have been used to predict
mobile-service QoS degradation from client-side measurements~\cite{kamel2011qos}.
Each parameter finally carries a physical reading for O-RAN allocation: $\mu_X$ is the
nominal packet gap of the encoded stream, $\sigma_X$ quantifies the jitter spread, and
$\xi$ indicates the tail regime.


\section{5G NR Physical Resource Block Capacity}
\label{sec:prb-analysis}
\label{sec:prb-capacity}

Effective PRB allocation for VR services depends on an accurate per-PRB capacity model. This
section derives the bit rate carried by one PRB from the governing 5G~NR parameters,
offering the analytical basis for the resource-allocation framework.

For a single PRB with $v$ spatial layers, the gross per-PRB throughput follows the 3GPP TS~38.306 peak-rate expression~\cite{3gpp_ts38306_v19}. With $N_{\mathrm{sc}}=12$
subcarriers per PRB, $N_{\mathrm{symb}}=14$ OFDM symbols per slot (normal cyclic prefix),
modulation order $Q_{\mathrm{m}}\in\{2,4,6,8\}$, code rate $R_c\le948/1024$ and
$N_{\mathrm{slots/s}}=2^{\mu}\times1000$ slots per second at numerology
$\mu\in\{0,1,2,3\}$~\cite{3gpp_ts38214_v19}, it reduces to

\begin{equation}
\begin{aligned}
R_{\mathrm{gross}}(\mu)&=v\,Q_{\mathrm{m}}\,R_c\,N_{\mathrm{sc}}\,N_{\mathrm{symb}}\,N_{\mathrm{slots/s}}\\
&=168{,}000\,v\,Q_{\mathrm{m}}\,R_c\,2^{\mu},
\end{aligned}
\label{eq:prb_throughput_mu}
\end{equation}
in bit/s per PRB, and the net rate subtracts the FR1 downlink control overhead,
\begin{equation}
R_{\mathrm{net}}=R_{\mathrm{gross}}\,(1-\mathrm{OH}),\qquad \mathrm{OH}=0.14 .
\label{eq:prb_throughput_net}
\end{equation}
The MCS, which the gNB selects from the CQI the UE reports, fixes $Q_{\mathrm{m}}$ and
$R_c$; the Transport Block Size then follows from the MCS, the PRB count and the OFDM
symbols~\cite{3gpp_ts38214_v19};
the Transport Block Size then follows from the MCS, the PRB count and the OFDM symbols~\cite{3gpp_ts38214_v19}. Table~\ref{tab:mcs_throughput}
evaluates \eqref{eq:prb_throughput_mu}--\eqref{eq:prb_throughput_net} for representative
MCS indices under the testbed configuration ($\mu=1$, $2\times2$ MIMO). Evaluated at the MCS implied by the reported CQI, and writing the per-RE spectral
efficiency as $\epsilon(\mathrm{CQI}_u)=Q_{\mathrm{m}}(\mathrm{CQI}_u)\,R_c(\mathrm{CQI}_u)$,
the net rate is the theoretical per-PRB efficiency used by the allocator (Tier~2,
Section~\ref{sec:rho-hierarchy}),
\begin{equation}
\rho_{\mathrm{th}}(\mathrm{CQI}_u)=168{,}000\,v\,\epsilon(\mathrm{CQI}_u)\,2^{\mu}(1-\mathrm{OH}),
\label{eq:net_per_prb_cqi}
\end{equation}

from which the minimum PRB count for UE~$u$ to meet its slice rate target $T(u)$ is
\begin{equation}
\mathrm{PRB}_u^{\min}=\left\lceil\frac{T(u)}{\rho_{\mathrm{th}}(\mathrm{CQI}_u)}\right\rceil .
\label{eq:prb_floor}
\end{equation}

\begin{table}[!t]
\caption{Per-PRB throughput for representative MCS indices ($\mu=1$, $2\times2$ MIMO,
$\mathrm{OH}=14\%$).}
\label{tab:mcs_throughput}

\centering
\small
\setlength{\tabcolsep}{4.5pt}
\begin{tabular}{cccccc}
\toprule
\textbf{MCS} & $\bm{Q_{\mathrm{m}}}$ & \textbf{Code rate} & \textbf{bits/RE} & \textbf{Gross} & \textbf{Net} \\
 & & $\bm{R_c}$ & & \textbf{Mbps/PRB} & \textbf{Mbps/PRB} \\
\midrule
0  & 2 & 0.1172 & 0.234 & 0.157 & 0.135 \\
9  & 2 & 0.6631 & 1.326 & 0.891 & 0.766 \\
16 & 4 & 0.6426 & 2.570 & 1.727 & 1.485 \\
25 & 6 & 0.8525 & 5.115 & 3.438 & 2.957 \\
29 & 8 & 0.7783 & 6.226 & 4.184 & 3.598 \\
31 & 8 & 0.9258 & 7.406 & 4.977 & 4.280 \\
\bottomrule
\end{tabular}
\end{table}

\subsection{Integration with QoE-Driven Resource Allocation}
\label{sec:prb-integration}
The PRB analysis links the physical-layer capacity model to the QoE framework. At runtime, the xApp evaluates the PRB floor~\eqref{eq:prb_floor} on the reported CQI to obtain the
slice's instantaneous requirement and the fallback estimate in the efficiency cascade of
Section~\ref{sec:rho-hierarchy}; combined with the multi-user traffic model of
Section~\ref{sec:multiuser-convolution} it also enables admission control, a new VR
connection being rejected when $\sum_{u=1}^{n}\mathrm{PRB}_u^{\min}>N_{\mathrm{PRB}}$.

The controller adopts the transport-level thresholds of IEEE~3079-2020~\cite{ieee3079} as
its compliance targets: jitter below 5~ms and packet loss below $1\%$, within an
end-to-end motion-to-photon budget of 20~ms at a frame rate of at least 90~Hz. Of these,
packet loss is the quantity most directly governed by PRB allocation, through the
RLC drop-rate mechanism detailed in Section~\ref{sec:results}.

\section{O-RAN Testbed Implementation}
\label{sec:testbed}

\subsection{System Architecture}
\begin{figure}[!t]
    \centering
    \includegraphics[width=\columnwidth]{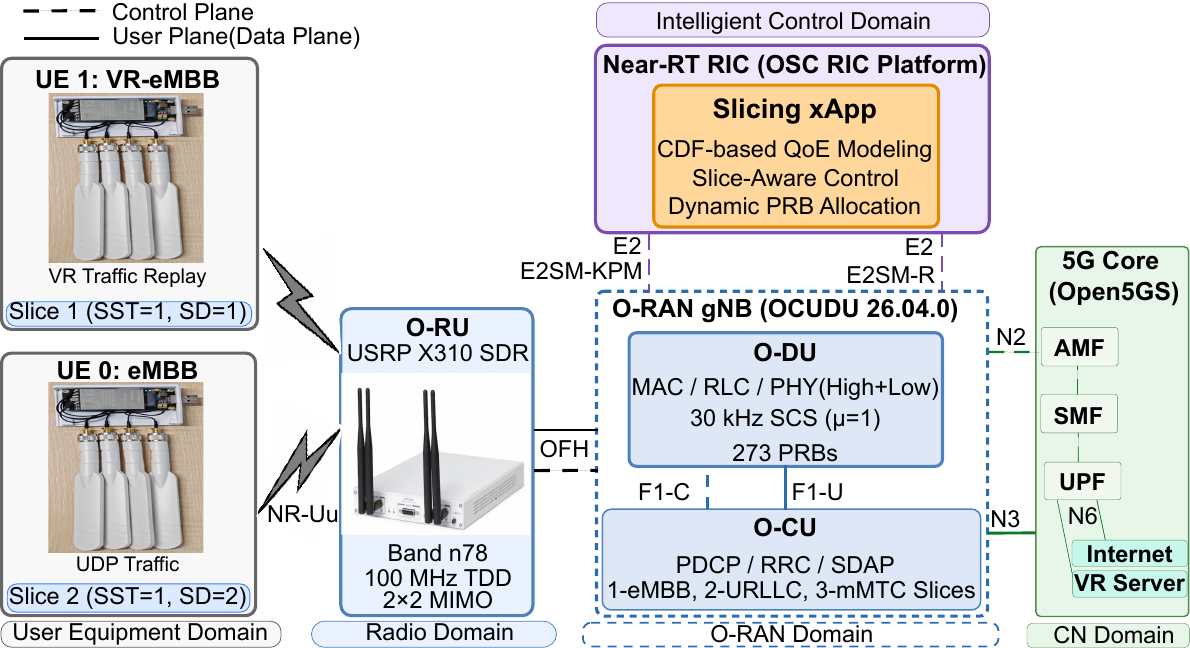}
    \vspace{-0.3cm}
    \caption{O-RAN testbed architecture.}
    \label{fig:oran}
    \vspace{-0.2cm}
\end{figure}

The testbed depicted in Fig.~\ref{fig:oran} builds a complete 5G O-RAN system: an Open5GS Core Network (CN),
an OCUDU~26.04 gNB, an OSC near-RT RIC running the custom Python
xApp, and two Quectel RMU~520 UEs served by a USRP~X310 (UBX-160) radio.

(SST${}={}$1) are differentiated by their Slice Differentiator: the priority \emph{VR-eMBB}
slice (SD${}={}$1, 250~Mbps DL target) carries the VR flow, and the general best-effort
\emph{eMBB} slice (SD${}={}$2, 100~Mbps DL target) carries the competing
load~\cite{3gpp.23.501.20}.

The radio link uses a 100~MHz channel in FR1 band~n78 (3.5~GHz) with numerology $\mu=1$
(30~kHz subcarrier spacing, normal cyclic prefix), giving $N_{\mathrm{PRB}}=273$ PRBs~\cite{3gpp_ts38101_1_v19}, and a $2\times2$ MIMO configuration enabling up to two layers and MCS indices $0$--$31$ (up to 256QAM). The 100~MHz bandwidth and 2T2R MIMO are the maxima the USRP~X310 supports in FR1, selected for peak rate. The $\mu=1$ numerology is a stability choice rather than a default: the 0.25~ms slot of 60~kHz spacing worsens the front end's occasional ``late'' and
underflow events into repeated radio-link failure, while 15~kHz spacing would cap the
usable bandwidth at 50~MHz. Measured per-PRB throughput agrees with
\eqref{eq:prb_throughput_mu}--\eqref{eq:prb_throughput_net} to within $0.5\%$ at MCS~31,
confirming the capacity model.
 
\subsection{E2SM-KPM Monitoring}
\label{sec:kpm-monitoring}
Four parallel E2SM-KPM Style~3 subscriptions are established maps at start-up and
reported by the OCUDU DU for all active UEs at its platform-minimum granularity of
1000~ms\footnote{The 1~Hz cadence is a platform limit, not a design choice: although the
near-RT RIC control window nominally spans 10~ms to 1~s, the OCUDU DU does not emit KPM
indications faster than once per second, so the control loop that consumes them runs at
1~Hz.}~\cite{3gpp-ts-28552}. 

\textbf{\texttt{DRB.UEThpDl}:} average per-UE downlink throughput, computed at
the gNB-DU from the RLC SDU volume successfully transmitted and acknowledged divided
by the transmission time (both excluding the final buffer-emptying slot). It serves as both a reallocation trigger
and the quality-score input.

\textbf{\texttt{CQI}:} the wideband CQI reported by the UEs, i.e.\ the index a
UE feeds back for the highest MCS it can decode at an acceptable error rate,
originating from the UE's own downlink channel-state measurements rather than a
gNB-side quantity. It is the Tier-2 efficiency fallback of
Section~\ref{sec:rho-hierarchy}.

\textbf{\texttt{DRB.AirIfDelayUl}:} arithmetic mean of the uplink air-interface
packet delay, a transport-block-level measurement with physical-layer start/stop
triggers~\cite{3gpp_ts_38_314_v19_0_0}. OCUDU~26.04 does not support the downlink
counterpart, so the uplink metric is used to monitor air-interface latency; it is recorded for compliance but, per Section~\ref{sec:triggers}, is not a trigger.

\textbf{\texttt{DRB.RlcPacketDropRateDl}:} the fraction of downlink RLC SDUs
discarded within the gNB-DU before any part is transmitted, i.e.\ congestion and
traffic-management drops inside the DU rather than radio losses after transmission.
It gates the IEEE~3079 loss budget for two reasons: 3GPP defines the drop rate as a
constituent of total packet loss~\cite{3gpp-ts-28552}, giving a conservative in-band lower
bound, and
congestion-induced DU drops are the loss mechanism most directly governed by PRB allocation.

\subsection{E2SM-RC Control}
\label{sec:e2sm-rc}
Control is released through O-RAN.WG3.E2SM-RC CONTROL Service Style~2, ``Radio Resource
Allocation Control'' (clause~7.6.3.1), Control Action ID~6, ``Slice-level PRB
quota''~\cite{oran_wg3_e2sm_rc_r004_v09_00}. In the OSC near-RT RIC this is called as \texttt{send\_control\_request\_style\_2\_action\_6(\dots, min\_prb\_ratio, max\_prb\_ratio,
dedicated\_prb\_ratio, \dots)}, whose three ratio arguments map onto the 3GPP
\texttt{RRMPolicyRatio} attributes \texttt{rRMPolicyMinRatio},
\texttt{rRMPolicyMaxRatio} and \texttt{rRMPolicyDedicatedRatio} of
TS~28.541~\cite{3gpp_ts_28_541_v20_2_0}, i.e.\ the prioritised, shared and dedicated
shares of radio resources for a slice instance. The control message includes the PLMN, the
S-NSSAI of the target slice, the gNB-DU-UE-F1AP-ID, and the three ratio parameters.


\section{Cross-Layer xApp Methodology}
\label{sec:xapp-methodology}

\begin{figure*}[t]
    \centering
    \includegraphics[width=0.8\textwidth]{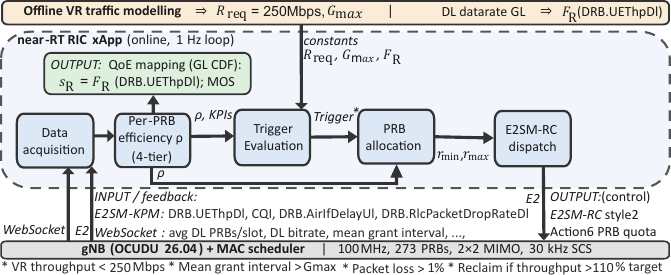}
    \caption{Closed-loop operation of the cross-layer QoE-driven slicing xApp.}
    \label{fig:xapp-blockdiagram}
\vspace{-0.4cm}    
\end{figure*}

The proposed controller is a single near-real-time RAN Intelligent Controller (near-RT
RIC) xApp that closes a 1~Hz loop between the application-layer statistics of the VR flow
and the physical-layer scheduling policy of the gNB. It has an
\emph{offline} stage, in which the fitted VR traffic distributions of
Sections~\ref{sec:traffic-char}--\ref{sec:multiuser-fit} are reduced to a small set of
control constants, and an \emph{online} stage that gathers per-UE telemetry, estimates
the achievable per-PRB efficiency, converts the downlink throughput to a reported quality
score, evaluates the reallocation triggers and dispatches a slice-level PRB quota over
E2SM-RC. Fig.~\ref{fig:xapp-blockdiagram} shows the complete data, control and feedback
paths.
 
\subsection{Offline-Derived Control Parameters}
\label{sec:offline-params}
\label{sec:gev-targets}
The fitted distributions of Sections~\ref{sec:traffic-char}--\ref{sec:multiuser-fit} are
reduced offline to the small set of constants the online loop uses. From the packet-IAT
GEV fit ($\xi$, $\mu_X$, $\sigma_X$), the fitted 99th-percentile of
the within-burst arrival timescale is $x_{0.99}\approx8.95~\mu$s, obtained from the standard GEV quantile inversion~\cite{coles2001extreme}. Three control quantities follow:

\textbf{Mean source rate and cross-validation.} The mean source rate
$\bar{R}_{\mathrm{src}}=8S_{\mathrm{pkt}}/\mathbb{E}[X]_{\mathrm{unc}}=151.5$~Mbps of
Section~\ref{sec:traffic-char} cross-validates the fits: because a data rate is, by
construction, $\mathrm{rate}\approx8S_{\mathrm{pkt}}/\mathrm{IAT}$, this must agree
with the mean of the independently fitted data-rate law, and the two routes agree to
within $3.1\%$.

\textbf{Slice service-rate target.} The slice is dimensioned for the \emph{peak} of the
empirical envelope rather than its mean, so the target is taken from the upper support of
the fitted \emph{data-rate} law, $R_{\mathrm{req}}=h_Y=250$~Mbps
(Table~\ref{tab:conv-constants}). This is consistent with the range reported in~\cite{3gppTR26928} for high-quality immersive services, roughly 100~Mbps for compressed delivery to
about 1~Gbps~\cite{3gppTR26928}, and serves as the primary throughput trigger.

\textbf{Scheduling-regularity requirement.} Essentially every arrival falls inside the slot already being served, so the GEV gives no packet-level service
deadline. What its upper tail establishes is a \emph{timescale separation}, placing
the regularity constraint one level up, on the \emph{grant} cadence; the resulting
bound $G_{\max}$ is derived in Section~\ref{sec:triggers}.

\subsection{Online Data Acquisition}
\label{sec:acquisition}
Two acquisition paths run in parallel and continuously; the efficiency-estimation
hierarchy of Section~\ref{sec:rho-hierarchy} chooses which feeds the allocator on any
given cycle. The first is the E2SM-KPM path of Section~\ref{sec:kpm-monitoring}, whose
four 1~Hz subscriptions provide the trigger, gate and quality-score inputs of
Fig~\ref{fig:xapp-blockdiagram}.

The second is gNB scheduler telemetry. Because the KPM counters reveal neither PRB occupancy nor MCS or HARQ statistics, the xApp opens an out-of-band WebSocket connection
(RFC~6455~\cite{rfc6455}) to the gNB's remote-control server, receiving the DU scheduler's
metrics as JSON at a 1~s interval. Of the reported fields, the controller uses the total
downlink PRBs scheduled, the MAC-layer downlink bitrate, the successful-HARQ count, the
per-window slot counts, and the OLLA-adjusted MCS, wideband CQI and rank indicator; the
remainder are collected only for per-UE link-health monitoring.

Let $p_{\mathrm{tot}}$ denote the total downlink PRBs the scheduler assigned to the UE
over the 1~s window, $N_{\mathrm{DL,slots}}$ the number of downlink slots in that window,
$R_{\mathrm{DL}}$ the MAC-layer downlink bitrate reported for the window, and
$n_{\mathrm{ok}}$ the number of successful downlink HARQ transmissions (first-pass ACKs)
in the window. On every ingest, the client derives
\begin{equation}
\begin{aligned}
\bar{p}_{\mathrm{DL}}&=\frac{p_{\mathrm{tot}}}{N_{\mathrm{DL,slots}}}, &\quad
\rho_{\mathrm{WS}}&=\frac{R_{\mathrm{DL}}}{\bar{p}_{\mathrm{DL}}},\\[3pt]
\Delta_{\mathrm{grant}}&=\frac{1000~\text{ms}}{n_{\mathrm{ok}}}, &\quad
\bar{B}_{\mathrm{TB}}&=\frac{R_{\mathrm{DL}}}{8\,n_{\mathrm{ok}}}.
\end{aligned}
\label{eq:ws-derived}
\end{equation}
Here $\bar{p}_{\mathrm{DL}}$ is the UE's steady-state PRB holding, the average number of
PRBs it is allocated in each downlink slot. The per-PRB efficiency $\rho_{\mathrm{WS}}$
is the downlink rate delivered per held PRB: $R_{\mathrm{DL}}$ is already a per-second
rate and $\bar{p}_{\mathrm{DL}}$ the corresponding steady-state allocation, so their
ratio is the throughput that one continuously scheduled PRB maintains. The mean
inter-grant interval $\Delta_{\mathrm{grant}}$ is not measured per grant but obtained as
the reciprocal grant rate over the window: the $n_{\mathrm{ok}}$ successful transport
blocks in 1~s, taken as approximately uniformly spaced, give a mean inter-grant time
of $1000/n_{\mathrm{ok}}$~ms (e.g.\ $n_{\mathrm{ok}}=1200$ ACKs gives
$\Delta_{\mathrm{grant}}\approx0.83$~ms). Finally, $\bar{B}_{\mathrm{TB}}$ is the mean
transport-block size in bytes.
 
\subsection{Radio-Parameter Engine: Per-PRB Efficiency Hierarchy}
\label{sec:rho-hierarchy}
On each control cycle the xApp selects the highest-fidelity available estimate of the
per-PRB efficiency $\rho$ (bit/s per PRB) through a three-tier cascade.
 
\textbf{Tier~0} (self-calibrating) calculates $\rho_0=R_{\mathrm{KPM}}/p_{\mathrm{prev}}$, the
measured \texttt{DRB.UEThpDl} divided by the PRB count corresponding to the previously
applied \emph{minimum} (guaranteed) quota; the guarantee is used rather than the ceiling
because a wide maximum quota that the UE only partly occupies would lower the ratio.
The estimate implicitly captures all MAC scheduling, HARQ, OLLA and header effects, but
is meaningful only when the UE is backlogged: an idle or lightly loaded UE leaves its
granted PRBs partly empty, so throughput divided by PRBs underestimates the true per-PRB
capacity. Tier~0 is therefore doubly gated. It is considered only when the measured throughput exceeds $20\%$ of the slice target, and the resulting estimate is accepted
only if it lies within $[50\%,120\%]$ of the Tier-2 theoretical per-PRB rate. At the
MCS~25 operating point of Table~\ref{tab:mcs_throughput}
($\rho_{\mathrm{th}}=2.957$~Mbps/PRB) the acceptance window is
$[1.479,\,3.548]$~Mbps/PRB, so a slice delivering 250~Mbps on 89 PRBs
($\rho_0=2.81$~Mbps/PRB) is accepted while a stale $4.2$~Mbps/PRB sample is rejected in
favour of the Tier-2 value. The lower bound rejects samples in which the granted quota
substantially exceeds the PRBs a bursty source actually fills; the upper bound rejects
the unrealistic values arising transiently when a throughput sample has not yet caught up
with a just-changed quota.

\textbf{Tier~1} (WebSocket-direct) sets $\rho_1=\rho_{\mathrm{WS}}$ from a single scheduler
window, so that numerator and denominator are mutually consistent and reflect the true OLLA
MCS and HARQ overhead; it requires WebSocket data fresher than 5~s and the same $20\%$ load
gate. \textbf{Tier~2} (theoretical) determines the per-PRB rate
$\rho_{\mathrm{th}}$ of~\eqref{eq:net_per_prb_cqi} from the effective MCS or the reported
CQI \cite{3gpp_ts38306_v19}, and is used as the cold-start and fallback value. In the reported runs Tier~1 is the
dominant source.
 
Algorithm~\ref{alg:xapp} outlines one control cycle, executed once per KPM indication
(1~Hz). Each cycle updates the ten-second rolling throughput average $\bar{R}_u$ (moving average) that
smooths the bursty VR traffic, evaluates the per-UE trigger condition of
Section~\ref{sec:triggers}, and, when at least one UE fires, runs the priority allocation
of Section~\ref{sec:allocation} before sending the resulting quotas. The MOS score in
line~1 is reported for evaluation only and never gates control.
 

\begin{algorithm}[t]
\footnotesize
\caption{QoE-driven slice PRB control (one near-RT cycle, 1~Hz).}
\label{alg:xapp}
\begin{algorithmic}[1]
\Require per-UE KPMs $\{R_u,\mathrm{CQI}_u,L_u,D_u\}$; scheduler telemetry
  $\{R^{\mathrm{ws}}_u,\bar p_u,\mathrm{MCS}_u,\Delta_u\}$; targets
  $T_{\mathrm{VR}}{=}250$, $T_{\mathrm{eMBB}}{=}100$~Mbps; PRB pool $N_{\mathrm{PRB}}$;
  over-provision $\omega{=}1.1$; gates $G_{\max},L_{\max}$; cooldowns
  $\tau_{\mathrm{VR}},\tau_{\mathrm{eMBB}}$; floors $\phi_{\mathrm{VR}},\phi_{\mathrm{eMBB}}$
\Statex \textit{Notation:} $\bar R_u$ 10-s avg.\ throughput; $\bar p_u$ mean PRBs per DL
  slot held by $u$ (the per-UE form of $\bar p_{\mathrm{DL}}$ in~\eqref{eq:ws-derived});
  $P_u$ last granted PRBs; $\pi(P){=}\lceil100P/N_{\mathrm{PRB}}\rceil$;
  $\mathrm{cap}_u$: over-provision ceiling $r_{\max}$ of~\eqref{eq:prb-ratios}
\Statex
\State \textbf{for all} $u$: update $\bar R_u$;\
  $\mathrm{MOS}_u \gets \mathrm{clip}(1{+}4F_R(R_u),1,5)$ \Comment{reported}
\State $C \gets \{\,u : \Call{Trigger}{u}\,\}$ \Comment{UEs to reallocate}
\State \textbf{if} $C=\varnothing$ \textbf{then return}
\State $\mathcal{A} \gets \Call{Allocate}{}$
\ForAll{$u \in C$}
  \State $(m_u,M_u) \gets \mathcal{A}[u]$;\
    \textbf{if} $P_u>0$: $M_u \gets \max\!\big(M_u,\lfloor 0.5\,\pi(P_u)\rfloor\big)$ \Comment{ramp guard}
  \State send E2SM-RC quota $(u,m_u,M_u)$;\ $P_u \gets m_u N_{\mathrm{PRB}}/100$;\
    $t^{c}_u \gets t$
\EndFor
\Statex
\Procedure{Trigger}{$u$}
  \State \textbf{if} $t^{c}_u=0$ \textbf{then return} \textsc{true} \Comment{reconnect}
  \State $\tau \gets \tau_{\mathrm{VR}}$ \textbf{if} $u\in\mathrm{VR}$ \textbf{else}
    $\tau_{\mathrm{eMBB}}$
  \State \textbf{if} $t-t^{c}_u<\tau$ \textbf{then return} \textsc{false} \Comment{cooldown}
  \If{$u \in \mathrm{VR}$}
    \State \textbf{if} $0<\bar R_u<T_{\mathrm{VR}}$ \textbf{then return} \textsc{true}
      \Comment{rate}
    \State \textbf{if} $R^{\mathrm{ws}}_u$ fresh $\wedge\ G_{\max}<\Delta_u<\infty
      \wedge\ \bar R_u>0.1\,T_{\mathrm{VR}}$ \textbf{then return} \textsc{true}
      \Comment{gap}
    \State \textbf{if} $L_u>L_{\max}$ \textbf{then return} \textsc{true} \Comment{loss}
  \EndIf
  \State \textbf{if} $\bar R_u \ge \omega\,T(u)$ \textbf{then return} \textsc{true}
    \Comment{reclaim (both slices)}
  \State \textbf{if} $t-t^{c}_u>30$~s \textbf{then return} \textsc{true} \Comment{periodic}
  \State \textbf{return} \textsc{false}
\EndProcedure
\Statex
\Procedure{Allocate}{}
  \ForAll{$u$} \Comment{tiered net bit/s per PRB}
    \If{$P_u{>}0 \wedge R_u{>}0.2\,T(u) \wedge \rho_0{\in}(0.5\rho_{\mathrm{th}},1.2\rho_{\mathrm{th}})$}
      \State $\rho_u \gets R_u/P_u$ \Comment{Tier 0: self-calibrated}
    \ElsIf{$R^{\mathrm{ws}}_u$ fresh}
      \State $\rho_u \gets R^{\mathrm{ws}}_u/\bar p_u$ \Comment{Tier 1: scheduler window}
    \Else
      \State $\rho_u \gets \rho_{\mathrm{th}}(\mathrm{MCS}_u,\mathrm{CQI}_u)$
        \Comment{Tier 2: \cite{3gpp_ts38306_v19}}
    \EndIf
    \State $k_u \gets \lceil T(u)/\rho_u \rceil$ \Comment{PRBs for target}
  \EndFor
  \State $P_{\mathrm{rem}} \gets N_{\mathrm{PRB}}$
  \State \textbf{for} $u\in\mathrm{VR}$: $a_u \gets \min(k_u,P_{\mathrm{rem}})$;\
    $P_{\mathrm{rem}} \mathrel{-}= a_u$ \Comment{VR-eMBB first}
  \State \textbf{for} $u\in\mathrm{eMBB}$: $a_u \gets \min(k_u,P_{\mathrm{rem}})$;\
    $P_{\mathrm{rem}} \mathrel{-}= a_u$ \Comment{then floor}
  \State $B_{\mathrm{rem}} \gets 100-\textstyle\sum_{u\in\mathrm{eMBB}}\phi_{\mathrm{eMBB}}$
    \Comment{reserve eMBB floors}
  \State \textbf{for} $u\in\mathrm{VR}$: $m_u \gets
    \mathrm{clip}(\pi(a_u),\phi_{\mathrm{VR}},B_{\mathrm{rem}})$;\
    $B_{\mathrm{rem}} \mathrel{-}= m_u$;\ $M_u \gets \mathrm{cap}_u$
  \State \textbf{for} $u\in\mathrm{eMBB}$: $m_u \gets
    \mathrm{clip}(\pi(a_u),\phi_{\mathrm{eMBB}},B_{\mathrm{rem}})$;\
    $B_{\mathrm{rem}} \mathrel{-}= m_u$;\ $M_u \gets \mathrm{cap}_u$
  \State \textbf{return} $\{(m_u,M_u)\}_u$ \Comment{min/max quota}
\EndProcedure
\end{algorithmic}

\end{algorithm}
 
\subsection{Quality Mapping via the Fitted Data-Rate CDF}
\label{sec:qoe-mapping}
QoS is measured through physically heterogeneous, dimensioned quantities, whereas QoE is
a bounded perceptual one, so a transformation onto a normalised scale $s\in[0,1]$ is
required. The CDF of an empirically fitted traffic model provides it directly: for a
variable $Z$ with $F_Z(z)=\Pr(Z\le z)$, the value $F_Z(z)$ is the fraction of the
modelled VR-traffic population experiencing a value no greater than $z$, so it ranks a
live measurement against the empirical distribution. It is also the appropriate mapping:
it is monotone, so a better metric never lowers satisfaction; bounded in $[0,1]$, so no
ad-hoc rescaling is needed~\cite{ross2019firstcourse}; and sigmoidal for the fitted
models, saturating to $0$ deep in the unacceptable region and to $1$ once conditions are
comfortably good.

Downlink throughput is a larger-is-better quantity modelled by the GL law of
Section~\ref{sec:Distribution-Fitting}. On every KPM indication the xApp therefore
computes the Type~I GL CDF~\cite{johnson1995continuous} at the raw
\texttt{DRB.UEThpDl} reading and maps it affinely to a Mean-Opinion-Score indicator,
\begin{equation}
\label{eq:qoe-mos}
\begin{gathered}
F_R(x)=\bigl(1+e^{-(x-\theta)/\sigma_Y}\bigr)^{-b},\\[2pt]
s_R=F_R\!\left(\texttt{DRB.UEThpDl}\right),\quad
\mathrm{MOS}=\mathrm{clip}\!\left(1+4s_R,1,5\right),
\end{gathered}
\end{equation}
with the shape, location and scale of Table~\ref{tab:conv-constants} ($b=1$, $F_R$
recovers the symmetric logistic). The affine map places it on the five-grade Absolute Category
Rating scale of ITU-T~P.800~\cite{itu_p800}, on which $5$ is excellent, $3$ fair and $1$
bad.

\begin{figure}[!t]
\centering
\includegraphics[width=0.86\columnwidth]{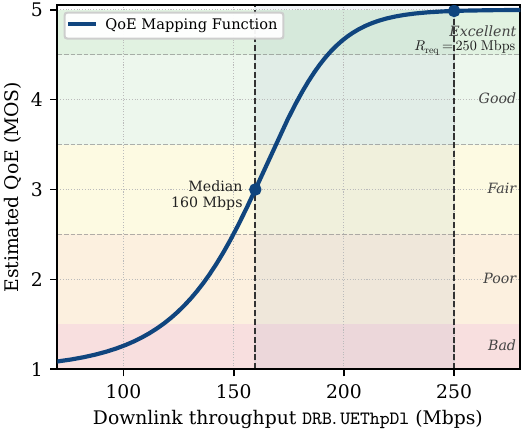}
\caption{Reported QoE mapping: the fitted Generalized Logistic CDF of the VR data rate maps
\texttt{DRB.UEThpDl} onto MOS.}
\label{fig:qoe-mapping}
\vspace{-0.2cm}
\end{figure}

Fig~\ref{fig:qoe-mapping} plots this mapping. The curve inherits the sigmoidal shape
of the CDF: it saturates towards $\mathrm{MOS}=5$ as the delivered rate approaches the
slice target $R_{\mathrm{req}}=250$~Mbps, rises steeply through the bulk of the
distribution, and decays towards $\mathrm{MOS}=1$ in the lower tail, where the RAN
delivers far less than the application demands. Because $F_R$ saturates as the
throughput approaches the target, the absolute value of this indicator is meaningful only when the offered load can actually reach that target; it is therefore used as a
relative, reported control-side signal in all runs, and its absolute value is reported
only for the sustained-rate case (Section~\ref{sec:results-qoe}).
 
The reference rate distinguishing satisfactory from unsatisfactory operation is the
\emph{median} of the fitted data-rate distribution ($159.8$~Mbps). Since $F_R$ compares the
delivered rate against the demand distribution, $F_R=0.5$ is by definition the throughput
at which half the demand instances are met, the natural boundary between under-served and
adequately-served operation; through~\eqref{eq:qoe-mos} it maps exactly to
$\mathrm{MOS}=3$, the ``fair'' acceptability midpoint, while a ``good'' grade
($\mathrm{MOS}=4$) corresponds to the 75th percentile ($\approx179$~Mbps). The mean is not
used: the fitted distribution is left-skewed, so the mean is shifted below the median
and maps to $\mathrm{MOS}\approx2.8$, whereas for a skewed quantity the median is the
robust measure of central tendency.

\subsection{Trigger Evaluation}
\label{sec:triggers}
An E2SM-RC reallocation is checked once per cycle per UE. For the VR-eMBB slice it
fires when any of the following hold: (i) the 10~s average throughput falls below the
$R_{\mathrm{req}}=250$~Mbps target; (ii) the mean inter-grant interval
$\Delta_{\mathrm{grant}}$ of~\eqref{eq:ws-derived} exceeds the scheduling-gap bound
$G_{\max}$, evaluated only when the WebSocket feed is fresher than 3~s and the slice
carries sustained traffic ($\bar R_u>0.1\,T_{\mathrm{VR}}$), so that inter-burst silence
cannot fire the gate; or (iii) the RLC drop rate exceeds the IEEE~3079 limit of $1\%$.
For either slice, an over-provision reclaim fires when the 10~s average throughput
reaches $\omega=110\%$ of its target in the reported runs; only the eMBB slice reaches
this ceiling, and a periodic re-evaluation fires every 30~s. A detected UE reconnection
forces an immediate re-quote, so that a new RNTI is not left with a stale full-bandwidth
allocation.

Each firing is subject to a per-slice inter-control cool-down (15~s for VR-eMBB, 5~s for
eMBB), several KPM reporting periods long, so that the windowed \texttt{DRB.UEThpDl}
average reflects the previous quota change before the next decision is taken: reacting
on every 1~s report would act on throughput that has not yet settled and can drive the
loop into oscillation, and the cool-down also limits the per-UE E2AP control-signalling
rate. The over-provision ratio $\omega=1.1$ was chosen empirically as the smallest
margin that reliably keeps the delivered eMBB throughput from falling below its target
after a reallocation.

Two points require emphasis. \emph{First}, the air-interface delay is \emph{not} a
trigger. On the USRP~X310 software-radio pipeline the measured
\texttt{DRB.AirIfDelayUl} has a fixed processing floor of approximately 7--8~ms that
is independent of radio quality and cannot be reduced by PRB reallocation; triggering on
it would fire the controller every cool-down interval with no possibility of
improvement. The violation is reported for compliance accounting but excluded from the
control law. \emph{Second}, the scheduling-gap bound $G_{\max}$ operates at the grant
timescale, not the packet timescale. Packet arrivals are a sub-slot quantity: the fitted
within-burst IAT has mean ${\approx}5.7~\mu$s and 99th percentile ${\approx}8.9~\mu$s,
and even the 99th percentile of the full arrival population ($344~\mu$s) lies below the
$500~\mu$s slot at $\mu=1$, so the slot-level scheduler groups hundreds of IP packets
into a single transport block per grant and gets nothing from issuing one grant per
packet arrival. $G_{\max}$ is therefore chosen as a small multiple of the slot period, 2~ms
(four $500~\mu$s slots) in this deployment, and the value is bracketed from both sides:
it must exceed the nominal inter-grant interval (${\approx}0.84$~ms in the reported
runs) so that the gate does not fire during healthy operation, and it must stay well
below the 8.33~ms frame period at 120~fps so that a sustained excess genuinely signals
that the slice is no longer served often enough to keep pace with frame-rate delivery.
At $G_{\max}=2$~ms, the slice is guaranteed at least four grants within each 8.33~ms
frame interval, the deadline by which the burst of packets carrying one frame
(${\approx}119$~kB, Section~\ref{sec:traffic-char}) must be cleared. Comparing the
WebSocket-reported $\Delta_{\mathrm{grant}}$ against $G_{\max}$ thus enforces scheduling
regularity, rather than instantaneous rate alone, at the cost of one comparison per
cycle and no online distribution evaluation.
 
\subsection{Priority-Based PRB Allocation and Control Dispatch}
\label{sec:allocation}
\label{sec:dispatch}
\label{sec:xlayer-summary}
When a trigger fires, the allocator converts the rate targets into integer PRB counts
using the current efficiency estimates, and bounds each slice
$s\in\{\mathrm{VR-eMBB},\mathrm{eMBB}\}$ at its over-provision ceiling:
\begin{equation}
N_{s}=\left\lceil\frac{T_{s}}{\rho_{s}}\right\rceil,
\qquad
N_{\mathrm{cap},s}=\left\lfloor\frac{\omega\,T_{s}}{\rho_{s}}\right\rfloor,
\qquad \omega=1.10 .
\label{eq:prb-counts}
\end{equation}
The allocator then divides the $N_{\mathrm{PRB}}=273$ PRBs of the cell by slice priority: $N_{\mathrm{VR}}$ PRBs are reserved for the VR-eMBB slice first, and the eMBB
slice is served from the remainder, so that under a genuine capacity shortage the
lower-priority eMBB slice yields rather than degrading VR-eMBB below its target. The
role of the ceiling differs by slice: for the eMBB slice it is what returns extra PRBs to the free pool once that slice exceeds $110\%$ of its target, while for the VR-eMBB
slice it merely bounds the grant just above the target that the priority allocation
$N_{\mathrm{VR}}$ already meets.

The absolute counts are mapped to the integer-percentage quotas carried by the E2SM-RC control message. The guaranteed share is rounded \emph{up}, so that quantisation never
costs a slice its target, while the ceiling is conservatively floored at the PRB-count
stage in~\eqref{eq:prb-counts}, so that the achievable rate never exceeds the $110\%$
cap:
\begin{equation}
r_{\min}=\left\lceil \frac{100\,N_{s}}{N_{\mathrm{PRB}}}\right\rceil\%,
\qquad
r_{\max}=\left\lfloor \frac{100\,N_{\mathrm{cap},s}}{N_{\mathrm{PRB}}}\right\rceil\% ,
\label{eq:prb-ratios}
\end{equation}
where $\lfloor\cdot\rceil$ denotes rounding to the nearest integer. Thus $r_{\min}$ is
the guaranteed share calculated from the slice's own rate target and measured efficiency,
and $r_{\max}$ is its over-provision ceiling. 

The computed quotas are sent through the E2SM-RC CONTROL Service Style~2, Control
Action~6 interface of Section~\ref{sec:e2sm-rc}, with the xApp setting the dedicated
ratio equal to $r_{\max}$. The gNB applies the quota to its MAC scheduler, which
enforces it per transmission-time interval, and the resulting change in delivered
throughput, PRB occupancy and grant interval is measured on the next KPM and WebSocket cycle, closing the loop of Fig.~\ref{fig:xapp-blockdiagram}.

\section{Experimental Results and Discussion}
\label{sec:results}
We evaluate the xApp across four use cases (UC1--UC4) on the hardware testbed of
Section~\ref{sec:testbed}. In every use case the priority slice (VR-eMBB), which carries the
VR flow at a 250~Mbps target, shares the cell with the eMBB slice (target 100~Mbps); the two
are isolated by the near-RT RIC only while the control xApp is active. UC1 calibrates the
isolated VR footprint. UC2 adds a well-behaved constant-bit-rate (CBR) neighbour on the eMBB
slice. UC3, the headline case, keeps the empirical VR replay but places it against a greedy
500~Mbps UDP subscriber on the eMBB slice. UC4 replaces the replay with a synthetic
fixed-rate 250~Mbps UDP source on the VR UE, again alongside the greedy 500~Mbps UDP
neighbour, to explore the worst-case, over-subscribed regime. 

 
 
\subsection{PRB Optimisation Framework}
\label{sec:results-prb}
 
\begin{figure}[t]
\centering
\includegraphics[width=0.9\columnwidth]{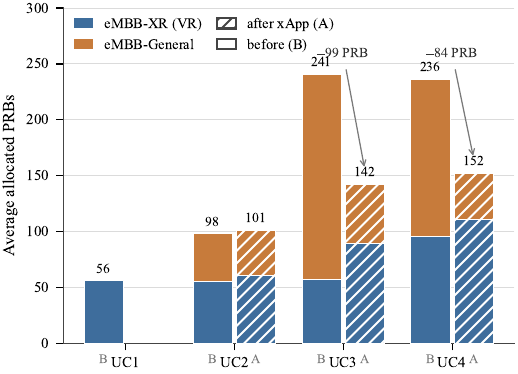}
\caption{Average PRB allocation per slice before (B) and after (A) xApp activation, UC1--UC4
(273-PRB cell).}
\label{fig:prb}
\vspace{-0.2cm}
\end{figure}
 
The central result is that the xApp recovers physical resources from an aggressive slice
only when real contention exists, and does so without starving the priority slice
(Fig.~\ref{fig:prb}). In UC2 the neighbouring eMBB slice is a disciplined CBR source
already fitting inside 43 PRBs, so there is nothing to reclaim: the total allocation is
essentially unchanged ($98\rightarrow101$ PRBs, a marginal

$3.1\%$ increase), and the xApp
merely rebalances a few PRBs toward the VR-eMBB slice ($55\rightarrow61$) while trimming the
eMBB slice ($43\rightarrow40$). This is the expected ``efficiency gap without contention'':
the value of dynamic slicing is latent until a slice misbehaves. That value becomes decisive in UC3. Under the scheduler default, the greedy 500~Mbps UDP flow holds 184 of the 273 PRBs and delivers 481~Mbps, crowding the VR-eMBB slice down to 57
PRBs. Activating the xApp inverts this: the E2SM-RC quota caps the greedy slice from 184 to
53 PRBs, a $71\%$ reduction on that slice, while simultaneously \emph{raising} the
protected VR-eMBB slice from 57 to 89 PRBs ($+56\%$) so that it can reach its service-rate
target. The net effect is a reduction of the total committed footprint from 241 to 142 PRBs,
i.e.\ \textbf{99 PRBs ($41.1\%$) returned to the free pool}, or equivalently $36.3\%$ of the
entire 273-PRB cell freed for admission of further slices. UC4 confirms the same mechanism under pure
over-subscription. The greedy UDP neighbour is cut from 140 to 41 PRBs ($-71\%$), and the
VR-eMBB slice is lifted from 96 to 111 PRBs to preserve priority. The total drops from 236
to 152 PRBs, so \textbf{84 PRBs ($35.6\%$) are freed}. In both contended cases the
slice-priority order is respected exactly: the aggressor yields by the same proportion
($71\%$), while the VR-eMBB slice is never reduced.
 
\subsection{Realistic Replay versus Synthetic Load}
\label{sec:results-realism}
\label{sec:results-traffic}
The architectural conclusion reached from these numbers is dependent on how the offered load is generated, and this is where the empirical traffic framework of
Sections~\ref{sec:traffic-char}--\ref{sec:multiuser-fit} changes the picture. UC1--UC3
drive the VR-eMBB slice with the VR traffic replay packets sent according to the fitted GEV inter-arrival and GL rate laws, reproducing the burst
structure of the source and the ${\sim}151$~Mbps mean of the fitted model. Whereas UC4
drives the same slice with a saturating 250~Mbps UDP source, the fixed-rate stand-in
commonly used in slicing studies. The two approaches place very different demands on the scheduler, and therefore lead to different capacity conclusions from identical
hardware.
 

The clearest evidence is the pre-xApp VR footprint. Under the realistic replay the VR-eMBB
slice occupies only 57 PRBs (UC3, before), whereas, under the constant 250~Mbps synthetic
load the nominally identical slice consumes 96 PRBs (UC4, before), a $68.4\%$ larger
footprint for the synthetic source. Equivalently, the empirical
replay requires just $59.4\%$ of the resources the fixed-rate load demands, so a study that
dimensioned this slice from a saturating fixed-rate assumption would over-provision it by
roughly two-thirds. The difference is a pure load effect, not a link artefact:
the per-PRB efficiency is essentially the same in both cases ${\sim}2.74$~Mbps/PRB at 156~Mbps for the replay versus ${\sim}2.67$~Mbps/PRB at 256~Mbps for the synthetic source.
Because both operate in the same MCS band (Table~\ref{tab:stability}), the synthetic source
simply asks for more bits, continuously, than a real VR session delivers on average.
Replaying empirically fitted traffic therefore reveals the true resource cost of a VR
slice, which fixed-rate testing materially overestimates, without sacrificing the
worst-case coverage that UC4 still provides.

\subsection{Testbed Validation and Stability}
\label{sec:results-stability}
Because the evaluation runs on real hardware, a USRP~X310 front end driving the OCUDU gNB
over the air rather than a link-abstracted simulation, the link-adaptation quantities
follow the instantaneous radio channel: the CQI, the MCS index and the selected modulation
order all vary as the propagation environment varies. Across all use cases the
reported CQI remains in the 14--15 band (mean 14.96) and the effective MCS index in the
24--25 band (mean 24.70), the physical layer operating mostly in 64-QAM with
occasional excursions to 256-QAM, as indicated in Table~\ref{tab:stability}.
 
\begin{table}[!t]
\caption{Link-level operating point per slice (mean $\pm$ std over each run).}
\label{tab:stability}
\centering
\footnotesize
\setlength{\tabcolsep}{5pt}
\begin{tabular}{clccc}
\toprule
 & & \multicolumn{2}{c}{\textbf{MCS index}} & \textbf{CQI}\\
\cmidrule(lr){3-4}
\textbf{UC} & \textbf{Slice} & Before & After & mean\\
\midrule
UC1 & VR-eMBB & $24.9\pm0.39$ & ---           & 15.0\\
UC2 & VR-eMBB & $25.0\pm0.62$ & $24.3\pm1.90$ & 15.0\\
UC2 & eMBB    & $25.5\pm0.68$ & $24.5\pm2.77$ & 15.0\\
UC3 & VR-eMBB & $24.9\pm0.46$ & $24.7\pm3.18$ & 15.0\\
UC3 & eMBB    & $24.0\pm0.92$ & $24.7\pm5.63$ & 14.9\\
UC4 & VR-eMBB & $24.4\pm0.87$ & $24.7\pm0.80$ & 14.9\\
UC4 & eMBB    & $24.9\pm0.33$ & $24.5\pm2.57$ & 15.0\\
\bottomrule
\end{tabular}
\vspace{-0.2cm}
\end{table}
 
Beyond the link-adaptation statistics, the stream-level delivery of the VR flow was measured independently of the per-UE scheduler counters. Under the control build, the VR RFC-3550 jitter was ${\sim}0.3$~ms and the delivery ratio was ${\sim}0.98$, even in the greedy-neighbour case UC3, where the unmanaged scheduler otherwise allows a best-effort UDP flow dominate. The priority slice therefore keeps low, stable latency while its resources are being reclaimed, confirming that the PRB reclamation of Section~\ref{sec:results-prb} is achieved without sacrificing VR responsiveness; these measures are reproduced across repetitions.
 
Despite these real-world variations, the environment is tightly bounded and statistically
stationary, which is what makes the cross-use-case PRB comparison solid. In every
baseline (scheduler-default) run the MCS standard deviation is below one index step
($0.33$--$0.92$) and the CQI standard deviation is negligible ($\le0.5$), i.e.\ the link
held a near-constant spectral efficiency for the full capture. Because the achievable bits
per PRB is thus effectively constant across use cases, any difference in PRB \emph{count}
between conditions is due to the offered load and to the xApp's allocation
policy rather than to opportunistic channel variation, so the differences in
Fig~\ref{fig:prb} are true policy effects. The larger MCS spread \emph{after} xApp
activation (up to $\pm5.6$ in the contended UC3) is not channel noise but the signature of
control: as the E2SM-RC quota repeatedly resizes each slice, the UE is driven across a
wider range of PRB grants and link-adaptation operating points within the same, unchanged
channel.

\subsection{Multi-User Capacity under Varying Channel Quality}
\label{sec:results-multiuser}
The preceding results describe the controller for a single VR user at a
near-constant operating point. The empirical traffic model of
Sections~\ref{sec:multiuser-convolution}--\ref{sec:multiuser-fit} and the per-PRB
capacity model of Section~\ref{sec:prb-analysis} together extend that behaviour
analytically to several concurrent VR users across the full channel-quality range,
answering the dimensioning question directly: given $n$ concurrent VR users and the
dominant channel quality, how many PRBs are required to maintain the target QoE?

The projection combines three results already established. First, the aggregate downlink
demand of $n$ statistically independent VR users is the $n$-fold convolution of the fitted
data-rate law~\eqref{eq:gl-nfold}; as the design statistic we take its $99$th percentile
$R_{\mathrm{agg}}(n)$, the rate that covers all but the heaviest $1\%$ of aggregate load,
which evaluates to $233$, $420$, $602$ and $779$~Mbps for $n=1$ to $4$. Second, the
wideband CQI reported by the UEs maps to a spectral efficiency through the 3GPP CQI table~\cite{3gpp_ts38214_v19}, and hence, through the net per-PRB
rate $\rho_{\mathrm{th}}(\mathrm{CQI})$ of~\eqref{eq:net_per_prb_cqi}, to the bit rate one
PRB delivers at that channel quality (Table~\ref{tab:mcs_throughput}). Third, substituting
$R_{\mathrm{agg}}(n)$ into the PRB-floor relation~\eqref{eq:prb_floor} yields the PRBs
required as a joint function of user count and channel quality,
\begin{equation}
N_{\mathrm{PRB}}^{\mathrm{req}}(n,\mathrm{CQI})=
\left\lceil \frac{R_{\mathrm{agg}}(n)}{\rho_{\mathrm{th}}(\mathrm{CQI})}\right\rceil .
\label{eq:prb-multiuser}
\end{equation}

\begin{figure}[!t]
\centering
\includegraphics[width=0.9\columnwidth]{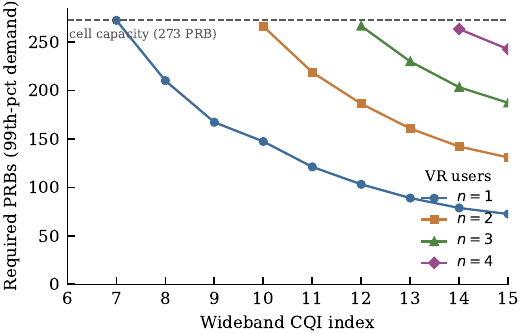}
\caption{Analytical PRBs required to sustain the $99$th-percentile aggregate demand of $n$
concurrent VR users, versus wideband CQI.}
\label{fig:prb-multiuser}
\vspace{-0.2cm}
\end{figure}

Fig.~\ref{fig:prb-multiuser} show this quantity over
the CQI range against the $N_{\mathrm{PRB}}=273$ PRBs of the cell. The two effects combine
as expected: the requirement increases with user count, because $R_{\mathrm{agg}}(n)$ grows
faster than linearly through the tail of the aggregate distribution, and it rises as the
channel worsen, because a lower CQI lowers $\rho_{\mathrm{th}}$ and each user's target
therefore costs more PRBs. Omitted entries are infeasible because the aggregate target cannot be
met within one cell, which is precisely the condition that the admission test of
Section~\ref{sec:prb-integration} rejects. At the cell edge of the reported deployment (CQI~$\approx$~10), a single VR
user is comfortably served with $148$ PRBs, two users require $267$ and sit at the limit,
and three or more must be admission-controlled; under strong channels (CQI~15) the cell supports up to four concurrent VR users within $243$ PRBs. The curve for each $n$
terminates where it crosses the capacity line, marking the lowest CQI at which that user
population can be admitted at the target QoE. Control overhead is already discounted through $(1-\mathrm{OH})$, but the table assumes the whole cell is available to the VR-eMBB slice.


A single-user measurement confirms the underlying trend: in the UC3 mobility captures
the xApp tracked the CQI variation with placement and kept the flow, the delivery
ratio holding at $0.996$, $0.992$ and $0.926$ across the three placements, the last with
elevated residual loss as the link approached the boundary that
Fig~\ref{fig:prb-multiuser} makes explicit. A full multi-user hardware test is
postponed to future work; the analytical model provides the envelope it would sample.

\subsection{Analytical and Perceptual QoE Estimation under Varying User Count and Channel Quality}
\label{sec:results-qoe-multiuser}
\label{sec:results-qoe}
The capacity envelope of Section~\ref{sec:results-multiuser} states \emph{whether} $n$
users can be served; the reported QoE model states \emph{how good} the experience is when
they are, and how it gets worse once they cannot. Because the throughput-satisfaction score
of~\eqref{eq:qoe-mos} is a per-user quantity, $\mathrm{MOS}=1+4\,F_R(x)$ with $x$ the rate the user actually receives, applying it to $(n,\mathrm{CQI})$ needs the delivered
per-user rate at each operating point. This follows from the priority policy of
Section~\ref{sec:allocation}. While the aggregate target is possible,
$N_{\mathrm{PRB}}^{\mathrm{req}}(n,\mathrm{CQI})\le N_{\mathrm{PRB}}$, every VR user is
served at its target $R_{\mathrm{req}}$ and $x=R_{\mathrm{req}}$. Once the cell saturates,
the priority scheduler provides the VR slice the whole cell, and the deliverable aggregate
$N_{\mathrm{PRB}}\,\rho_{\mathrm{th}}(\mathrm{CQI})$ is shared across the $n$ users, so each
receives
\begin{equation}
x(n,\mathrm{CQI})=
\begin{cases}
R_{\mathrm{req}}, & n\,R_{\mathrm{req}}\le N_{\mathrm{PRB}}\,\rho_{\mathrm{th}}(\mathrm{CQI}),\\[4pt]
\dfrac{N_{\mathrm{PRB}}\,\rho_{\mathrm{th}}(\mathrm{CQI})}{n}, & \text{otherwise},
\end{cases}
\label{eq:per-user-rate}
\end{equation}
and its QoE is $\mathrm{MOS}(n,\mathrm{CQI})=1+4\,F_R\!\big(x(n,\mathrm{CQI})\big)$. The
equal-share form in the saturated branch is the graceful-degradation assumption; the
alternative, in which admission control simply refuses the excess users, would hold the
admitted users at $\mathrm{MOS}\approx5$ and is the binary counterpart of the same
boundary.

\begin{figure}[!t]
\centering
\includegraphics[width=0.9\columnwidth]{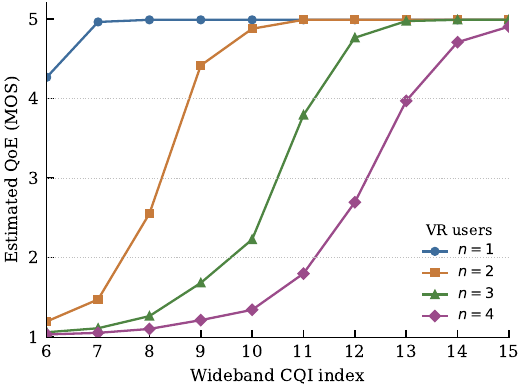}
\caption{Analytical per-user QoE (MOS) for $n$ concurrent VR users versus wideband CQI.}
\label{fig:qoe-multiuser}
\vspace{-0.2cm}
\end{figure}

Fig~\ref{fig:qoe-multiuser} shows this mapping over the CQI range. Two regimes are
visible. Where the cell meets the aggregate target the delivered rate equals
$R_{\mathrm{req}}$ and, since $F_R(R_{\mathrm{req}})\approx0.997$, the QoE sits at the
ceiling $\mathrm{MOS}\approx5$ regardless of $n$; every curve is flat and excellent in this
region. As the channel degrades each curve reaches a knee at the CQI where its user
population first saturates the cell, after which QoE drops steeply: at CQI~10, for
instance, the MOS is $5.0$, $4.9$, $2.2$ and $1.3$ for $n=1$ to $4$, as the per-user share
drops to $250$, $215$, $144$ and $108$~Mbps. The knee moves to higher CQI as $n$ grows,
because a larger population saturates the cell at a milder channel deficit. The two
figures are complementary views of one boundary Fig.~\ref{fig:prb-multiuser} locates it
in PRBs, Fig.~\ref{fig:qoe-multiuser} gives the perceptual cost of crossing it, and together they let an operator identify, for a given user count and channel quality, both
the PRB budget and the QoE the VR slice will sustain. The sharpness of the knee should be interpreted as a property of the model, not as a prediction of the transient. It has two sources. First, the mapping saturates because $F_R(R_{\mathrm{req}})\approx0.997$ leaves virtually no headroom above the target; thus, feasible points are pinned at the ceiling, compressing the MOS dynamic range between saturation and the cell limit. Second, the feasibility test is a step: the delivered rate in~\eqref{eq:per-user-rate} switches from $R_{\mathrm{req}}$ to the equal share at a single CQI, with no intermediate regime. A real scheduler crosses this boundary more gradually because link adaptation, HARQ retransmissions, and the finite cooldown of Section~\ref{sec:triggers} spread the transition over several control cycles, while partial slice service degrades progressively rather than switching state. The curves should therefore be interpreted as bounding the operating envelope and locating the CQI at which degradation begins, rather than as the time-domain trajectory through that boundary.

This analytical surface is based empirically at its single measured point. Because
$F_R$ saturates near the target, the absolute MOS is meaningful only when the offered load
can reach it, which on the testbed holds for UC4 alone: there the slice delivers
256~Mbps, maps to $\mathrm{MOS}\approx5.0$, and \emph{retains} it under xApp control even
as the greedy neighbour is limited from 140 to 41 PRBs, so priority is preserved
perceptually and not merely in PRB count, exactly the feasible-region ceiling of the
$n=1$ curve in Fig.~\ref{fig:qoe-multiuser}. In UC1--UC3 the replay's empirical mean
(${\sim}151$~Mbps) sits near the median of $F_R$, so the indicator serves only as a
relative control-side signal and VR quality is measured through the objective delivery-ratio, loss and latency measures of Section~\ref{sec:results-stability}.

\section{Conclusion}
\label{sec:conclusion}
This paper introduced a QoE-driven O-RAN network-slicing framework for Virtual Reality
whose quality model and control policy are based on empirically fitted traffic
statistics. From 38.6 million captured packets we fitted a Generalized Extreme Value law
to the inter-arrival times and a Generalized Logistic law to the downlink data rates, and
derived closed-form $n$-fold convolutions for the aggregate traffic of $n$ users,
validated across one to four. The rate CDF serves as a percentile-based satisfaction
indicator, the inter-arrival law sets the slice targets offline, and a near-real-time RIC
xApp closes the loop over E2SM-KPM and E2SM-RC.

On the OCUDU testbed, the controller reclaimed up to $41\%$ of the committed PRBs from a
greedy neighbour without reducing the protected VR-eMBB slice, and held packet loss to
$0.037\%$, within the IEEE~3079-2020 budget. A controlled comparison showed a fixed-rate
synthetic source overstates the VR slice's PRB footprint by $68\%$, an effect invisible to
conventional evaluation; an analytical study then located the admission boundary across
user counts and channel qualities, and the graceful QoE degradation beyond it.

Because every online signal traverses standardised interfaces, the controller is portable
to any O-RAN-compliant gNB exposing the required KPMs. Future work will test the analytical
capacity envelope with a concurrent-UE hardware campaign, extend it to multi-cell and
handover cases and to AR, MR and cloud VR, and replace the SDR-limited delay floor with a
direct latency measurement.
 
 \section*{Acknowledgment}
 This research was supported by UKRI through the EPSRC under grants: the Technology Missions Fund project CHEDDAR (EP/Y037421/1). AI was used solely to improve the grammar and readability of the manuscript. The authors are fully responsible for all scientific content.
 
\bibliographystyle{IEEEtran}
\bibliography{references}

@IEEEtranBSTCTL{IEEEexample:BSTcontrol,
  CTLdash_repeated_names = "no"
}

@inproceedings{zhao2021virtual,
  title={Virtual reality gaming on the cloud: A reality check},
  author={Zhao, Sihao and Abou-zeid, Hatem and Atawia, Ramy and Manjunath, Yoga Suhas Kuruba and Sediq, Akram Bin and Zhang, Xiao-Ping},
  booktitle={2021 IEEE Global Communications Conference (GLOBECOM)},
  pages={1--6},
  year={2021},
  organization={IEEE}
}

@techreport{3gppTR26928,
  author      = {{3GPP}},
  title       = {{Group Services and System Aspects; Extended Reality (XR) in 5G }},
  type        = {Technical Specification (TS)},
  institution = {{3rd Generation Partnership Project}},
  number      = {26.928},
  version     = {V18.0.0},         
  note        = {Release 18},
  year        = {2023},           
  month       = {03},             
  url         = {https://www.3gpp.org/ftp/Specs/archive/26_series/26.928/26928-i00.zip}
}

@techreport{abberger2002ml,
  title={ML-estimation in the location-scale-shape model of the generalized logistic distribution},
  author={Abberger, Klaus},
  year={2002},
  institution={CoFE Discussion Paper}
}

@inproceedings{chyad2025modelling,
  title={Modelling 6Dof Vr Gaming Traffic for Next-Generation Networks},
  author={Chyad, Ahmed and Shah, Syed Danial Ali and Qazzaz, Mohammed MH and Hafeez, Maryam and Zaidi, Syed Ali Raza},
  booktitle={2025 Joint European Conference on Networks and Communications \& 6G Summit (EuCNC/6G Summit)},
  pages={157--162},
  year={2025},
  organization={IEEE}
}

@techreport{3gpp.23.501.20,
  author      = {{3GPP}},
  title       = {System architecture for the 5G System (5GS); Stage 2},
  type        = {Technical Specification (TS)},
  number      = {23.501},
  institution = {{3rd Generation Partnership Project}},
  year        = {2025},
  month       = {dec},
  note        = {Version 20.0.0, Release 20},
  url         = {https://www.3gpp.org/DynaReport/23501.htm}
}

@article{KullbackLeibler1951,
  author  = {Kullback, S. and Leibler, R. A.},
  title   = {On Information and Sufficiency},
  journal = {The Annals of Mathematical Statistics},
  year    = {1951},
  volume  = {22},
  number  = {1},
  pages   = {79--86},
  doi     = {10.1214/aoms/1177729694}
}

@book{kotz2000extreme,
  title={Extreme value distributions: theory and applications},
  author={Kotz, Samuel and Nadarajah, Saralees},
  year={2000},
  publisher={world scientific}
}

@techreport{3gpp_ts38214_v19,
  title        = {{NR; Physical layer procedures for data}},
  author       = {{3GPP}},
  institution  = {3rd Generation Partnership Project},
  number       = {TS 38.214 V19.2.0},
  month        = feb,
  year         = {2026},
  note         = {Sec5.1.3 (MCS tables); Table 5.1.3.1-2 (256QAM MCS index table);
                  Sec5.1.3.2 (TBS procedure); Sec5.2.2 (CSI feedback);
                  Sec5.2.2.1 and Table 5.2.2.1-2 (CQI reporting for 256QAM UEs)}
}

@techreport{3gpp_ts38306_v19,
  title        = {{NR; User Equipment (UE) radio access capabilities}},
  author       = {{3GPP}},
  institution  = {3rd Generation Partnership Project},
  number       = {TS 38.306 V19.1.0},
  month        = feb,
  year         = {2026},
  note         = {Section 4.1.2: Peak data rate calculation formula, scaling factor f, overhead OH}
}

@techreport{3gpp_ts38101_1_v19,
  title        = {{NR; User Equipment (UE) radio transmission and reception;
                  Part 1: Range 1 Standalone}},
  author       = {{3GPP}},
  institution  = {3rd Generation Partnership Project},
  number       = {TS 38.101-1 V19.4.0},
  month        = mar,
  year         = {2026},
  note         = {Table 5.3.2-1: NR FR1 channel bandwidth and PRB allocation per numerology}
}

@inproceedings{paymard2023performance,
  title={Performance of joint XR and best-effort eMBB traffic in 5G-advanced networks},
  author={Paymard, Pouria and Amiri, Abolfazl and Kolding, Troels E and Pedersen, Klaus I},
  booktitle={2023 IEEE 97th Vehicular Technology Conference (VTC2023-Spring)},
  pages={1--5},
  year={2023},
  organization={IEEE}
}

@article{fisher1928limiting,
  author    = {R. A. Fisher and L. H. C. Tippett},
  title     = {Limiting Forms of the Frequency Distribution of the Largest or Smallest Member of a Sample},
  journal   = {Mathematical Proceedings of the Cambridge Philosophical Society},
  volume    = {24},
  number    = {2},
  pages     = {180--190},
  month     = apr,
  year      = {1928},
  doi       = {10.1017/S0305004100015681}
}

@book{coles2001extreme,
  author    = {Stuart Coles},
  title     = {An Introduction to Statistical Modeling of Extreme Values},
  publisher = {Springer},
  address   = {London, UK},
  year      = {2001},
  isbn      = {9781852334598}
}

@inproceedings{liu2007evtwireless,
  author    = {C. Liu and Y. Shu and J. Liu and O. W. W. Yang},
  title     = {Application of Extreme Value Theory to the Analysis of Wireless Network Traffic},
  booktitle = {Proceedings of the IEEE International Conference on Communications (ICC)},
  address   = {Glasgow, UK},
  month     = jun,
  year      = {2007},
  pages     = {486--491},
  doi       = {10.1109/ICC.2007.85}
}

@techreport{3gpp38838,
  author      = {{3GPP}},
  title       = {Study on XR (Extended Reality) Evaluations for NR},
  institution = {3rd Generation Partnership Project},
  type        = {Technical Report},
  number      = {TR 38.838 V17.0.0},
  month       = jan,
  year        = {2022}
}

@techreport{3gpp26926,
  author      = {{3GPP}},
  title       = {Traffic Models and Quality Evaluation Methods for Media and XR Services in 5G Systems},
  institution = {3rd Generation Partnership Project},
  type        = {Technical Report},
  number      = {TR 26.926 V18.2.0},
  month       = mar,
  year        = {2024}
}

@inproceedings{kamel2011qos,
  author    = {A. Kamel and A. Al-Fuqaha},
  title     = {Client-Side Architecture for Mobile Service QoS Monitoring Using Generalized Extreme Value Theorem},
  booktitle = {Proceedings of the IEEE GLOBECOM Workshops},
  address   = {Houston, TX, USA},
  month     = dec,
  year      = {2011},
  doi       = {10.1109/GLOCOMW.2011.6162384}
}

@book{rinne2008weibull,
  author    = {Horst Rinne},
  title     = {The Weibull Distribution: A Handbook},
  publisher = {CRC Press},
  year      = {2008},
  isbn      = {9781420087444}
}

@book{zolotarev1986stable,
  author    = {V. M. Zolotarev},
  title     = {One-Dimensional Stable Distributions},
  publisher = {American Mathematical Society},
  address   = {Providence, RI, USA},
  year      = {1986},
  isbn      = {9780821895030}
}

@standard{ieee3079,
  title        = {IEEE Standard for Head-Mounted Display (HMD)-Based Virtual Reality (VR) Sickness Reduction Technology},
  organization = {IEEE},
  number       = {IEEE Std 3079-2020},
  address      = {New York, NY, USA},
  year         = {2020}
}

@inproceedings{kumar2021qoe,
  author    = {A. Kumar and M. Abhishek and V. Muthukumar},
  title     = {{QoE} Prediction for {VR} Video Streaming Using Deep Learning},
  booktitle = {Proceedings of the IEEE International Conference on Communications (ICC)},
  address   = {Montreal, QC, Canada},
  pages     = {1--6},
  year      = {2021},
  doi       = {10.1109/ICC42927.2021.9500804}
}

@article{zhang2017network,
  author    = {S. Zhang},
  title     = {An Overview of Network Slicing for 5G},
  journal   = {IEEE Wireless Communications},
  volume    = {26},
  number    = {3},
  pages     = {111--117},
  month     = jun,
  year      = {2019},
  doi       = {10.1109/MWC.2019.1800234}
}

@article{condoluci2017enabling,
  author    = {M. Condoluci and T. Mahmoodi},
  title     = {Softwarization and Virtualization in 5G Mobile Networks: Benefits, Trends and Challenges},
  journal   = {Computer Networks},
  volume    = {146},
  pages     = {65--84},
  month     = dec,
  year      = {2018},
  doi       = {10.1016/j.comnet.2018.09.005}
}

@techreport{3gpp-ts-28552,
  author      = {{3GPP}},
  title       = {{Management and orchestration; 5G performance measurements}},
  type        = {Technical Specification (TS)},
  institution = {{3rd Generation Partnership Project}},
  number      = {28.552},
  version     = {20.2.0},         
  note        = {Release 20},
  year        = {2026},           
  month       = {03},             
  url         = {https://www.3gpp.org/ftp/Specs/archive/28_series/28.552/}
}

@techreport{3gpp_ts_38_314_v19_0_0,
  author       = {{3GPP}},
  title        = {{NR; Layer 2 Measurements}},
  type        = {Technical Specification (TS)},
  institution  = {{3rd Generation Partnership Project}},
  number       = {TS 38.314 V19.0.0},
  year         = {2025},
  month        = 09,

}

@techreport{3gpp_ts_28_541_v20_2_0,
  author       = {{3GPP}},
  title        = {{Technical Specification Group Services and System Aspects; 
                   Management and orchestration; 
                   5G Network Resource Model (NRM); 
                   Stage 2 and stage 3 (Release 20)}},
  institution  = {3GPP},
  number       = {TS 28.541 V20.2.0},
  year         = {2026},
  month        = mar,
  type         = {Technical Specification},
}

@techreport{oran_wg3_e2sm_rc_r004_v09_00,
  author       = {{O-RAN Alliance WG3}},
  title        = {{Near-Real-time RAN Intelligent Controller; 
                   E2 Service Model (E2SM), RAN Control}},
  institution  = {O-RAN Alliance},
  type         = {Technical Specification},
  number       = {O-RAN.WG3.E2SM-RC-R004-v09.00},
  year         = {2025},
}

@book{ross2019firstcourse,
  author    = {Sheldon Ross},
  title     = {A First Course in Probability},
  edition   = {10},
  publisher = {Pearson},
  year      = {2019}
}

@inbook{johnson1995continuous,
  author    = {Johnson, N. L. and Kotz, S. and Balakrishnan, N.},
  title     = {Continuous Univariate Distributions},
  volume    = {2},
  edition   = {2nd},
  chapter   = {23},
  year      = {1995},
  publisher = {Wiley},
  address   = {New York}
}

@techreport{rfc6455,
  author      = {I. Fette and A. Melnikov},
  title       = {The WebSocket Protocol},
  number      = {6455},
  institution = {RFC Editor},
  type        = {Request for Comments},
  year        = {2011},
  month       = dec,
  doi         = {10.17487/RFC6455},
  url         = {https://www.rfc-editor.org/info/rfc6455}
}

@techreport{itu_p800,
    author      = {{ITU-T}},
    title       = {Methods for Subjective Determination of Transmission Quality},
    institution = {International Telecommunication Union},
    type        = {Recommendation},
    number      = {ITU-T P.800},
    year        = {1996},
    }

@book{feller1966introduction,
  title={An Introduction to Probability Theory and Its Applications},
  author={Feller, William},
  volume={2},
  year={1966},
  publisher={John Wiley \& Sons}
}

@article{korneev2024cloudvr,
  author={Korneev, Eugene and Liubogoshchev, Mikhail and Bankov, Dmitry and Khorov, Evgeny},
  title={How to Model Cloud {VR}: An Empirical Study of Features That Matter},
  journal={IEEE Open Journal of the Communications Society},
  volume={5}, pages={4155--4170},
  year={2024}}

@article{liubogoshchev2021adaptive,
  author={Liubogoshchev, Mikhail and Ragimova, Kamila and Lyakhov, Andrey and Tang, Steffen and Khorov, Evgeny},
  title={Adaptive Cloud-Based Extended Reality: Modeling and Optimization},
  journal={IEEE Access}, volume={9}, pages={35287--35299}, year={2021}}

@article{ruan2021survey,
  author={Ruan, Jinjia and Xie, Dongliang},
  title={A Survey on {QoE}-Oriented {VR} Video Streaming: Some Research Issues and Challenges},
  journal={Electronics}, volume={10}, number={17}, pages={2155}, year={2021}}

@article{chiariotti2021survey,
  author={Chiariotti, Federico},
  title={A Survey on 360-Degree Video: Coding, Quality of Experience and Streaming},
  journal={Computer Communications}, volume={177}, pages={133--155}, year={2021}}

@article{kougioumtzidis2023deep,
  author={Kougioumtzidis, Georgios and Vlahov, Atanas and Poulkov, Vladimir K. and Lazaridis, Pavlos I. and Zaharis, Zaharias D.},
  title={Deep Learning-Aided {QoE} Prediction for Virtual Reality Applications Over Open Radio Access Networks},
  journal={IEEE Access}, volume={11}, pages={143514--143529}, year={2023}}

@inproceedings{casparsen2024near,
  author={Casparsen, Andreas and Soret, Beatriz and Nielsen, Jimmy Jessen and Popovski, Petar},
  title={Near Real-Time Data-Driven Control of Virtual Reality Traffic in Open Radio Access Network},
  year={2023},
  booktitle={2023 IEEE Global Communications Conference (GLOBECOM)},
  pages={3481
--3486},
  organization={IEEE}
}


\end{document}